# Query-Monotonic Turing Reductions[*]

Lane A. Hemaspaandra[†]
Department of Computer Science
University of Rochester
Rochester, NY 14627, USA

Mayur Thakur[‡]
Department of Computer Science
University of Missouri–Rolla
Rolla, MO 65409, USA

January 31, 2006

## Abstract

We study reductions that limit the extreme adaptivity of Turing reductions. In particular, we study reductions that make a rapid, structured progression through the set to which they are reducing: Each query is strictly longer (shorter) than the previous one. We call these reductions query-increasing (query-decreasing) Turing reductions. We also study query-nonincreasing (query-nondecreasing) Turing reductions. These are Turing reductions in which the sequence of query lengths is nonincreasing (nondecreasing). We ask whether these restrictions in fact limit the power of reductions. We prove that query-increasing and query-decreasing Turing reductions are incomparable with (that is, are neither strictly stronger than nor strictly weaker than) truth-table reductions and are strictly weaker than Turing reductions. In addition, we prove that query-nonincreasing and query-nondecreasing Turing reductions are strictly stronger than truth-table reductions and strictly weaker than Turing reductions. Despite the fact that we prove query-increasing and query-decreasing Turing reductions to in the general case be strictly weaker than Turing reductions, we identify a broad class of sets $A$ for which any set that Turing reduces to $A$ will also reduce to $A$ via both query-increasing and query-decreasing Turing reductions. In particular, this holds for all *tight paddable* sets, where a set is said to be tight paddable exactly if it is paddable via a function whose output length is bounded tightly both from above and from below in the length of the input. We prove that many natural NP-complete problems such as satisfiability, clique, and vertex cover are tight paddable.

## 1 Introduction

Oracle access is an important notion in the theory of computation. It forms the basis for defining the different levels of the polynomial hierarchy [MS72,Sto76]. It is central to the notion of Turing reducibility [LLS75], which is used to compare the complexity of problems.

[*]Supported in part by grant NSF-CCF-0426761. A preliminary version of this paper was presented at the Eleventh Annual International Computing and Combinatorics Conference (COCOON 2005) [HT05].
[†]URL: http://www.cs.rochester.edu/u/lane.
[‡]URL: http://web.umr.edu/~thakurk. Work done in part while at the University of Rochester.

How the *nature* of access to its oracle affects the power of a Turing machine is a central research issue. For example, the issue of whether adaptive access is more powerful than nonadaptive access has been well researched (see, e.g., [Wat87,AB00,PS02]).[1] A Turing machine with adaptive access to its oracle has the full flexibility of asking queries in any order. In particular, the $k$th query may depend on the answers to the previous $k-1$ queries. On the other hand, in nonadaptive access, the Turing machine is required to generate all queries before asking any of them. Thus, adaptive access and nonadaptive access might at first seem to represent two extremes of flexibility in deterministic oracle access (when one is without external limits on the number of queries allowed).

In this paper, we examine modes of oracle access whose flexibility in some cases lies somewhere between these two extremes, and in some cases is incomparable even to the lower "extreme," truth-table reducibility. To help us formalize the degree and nature of the flexibility of oracle access, we introduce the notion of *query-restricted* Turing reductions, $\leq^p_{\rho\text{-}T}$, where $\rho$ is a set of query sequences. A Turing machine $M$ has the $\rho$ query property with respect to $B$ if the following holds for each input $x$: If $q_1, q_2, \ldots, q_k$ is the sequence of strings queried by $M^B(x)$, then $(x, q_1, q_2, \ldots, q_k) \in \rho$. We say that $A \leq^p_{\rho\text{-}T} B$ if there is a deterministic Turing machine $M$ such that $M$ robustly (i.e., for all oracles) runs in polynomial time, $L(M^B) = A$, and $M$ has the $\rho$ query property with respect to $B$.

Note that each query-restricted Turing reduction $\leq^p_{\rho\text{-}T}$ imposes certain restrictions (formally captured by $\rho$) on the sequence of strings queried by the machine. In this paper, we study query-restricted Turing reductions in which the set of allowable query sequences imposes monotonicity in the *length* of the queries that the underlying machine asks to the oracle. We call such reductions *query-monotonic Turing reductions*. For example, in *query-increasing* Turing reductions, the machine is required to ask its queries in a length-increasing fashion, i.e., each query must be longer than all the earlier ones (and also, in its strongest form, longer than the input). The main query-monotonic reductions that we study in this paper are query-increasing reductions ($\leq^p_{li\text{-}T}$), query-decreasing reductions ($\leq^p_{ld\text{-}T}$), query-nonincreasing reductions ($\leq^p_{lni\text{-}T}$), and query-nondecreasing reductions ($\leq^p_{lnd\text{-}T}$). (Formal definitions of these and other query-monotonic reductions are provided in Section 2.)

We came to define the notion of query-monotonicity motivated by the idea of rapid progress. In particular, in this paper we wish to study the power of polynomial-time machines that sweep, directionally, through their database (i.e., oracle). Such machines will have a "use it when you can" flavor to their access to the information at each length—once one query at or beyond a length is asked, the rest of the information at that length is forever lost to direct access, on the current input. However, this apparently restrictive access is not necessarily restrictive in effect. It is at least plausible that, by exploiting properties of particular databases, the restriction (for those databases) will be toothless, i.e., can be obeyed without loss of generality. In this paper, we show that in some cases the restriction has teeth, but we also show that in other quite central cases the restriction is toothless. In particular, for several reductions $\leq^p_\alpha$ and $\leq^p_\beta$, we ask whether, for all $A, B \subseteq \Sigma^*$, $A \leq^p_\alpha B$ implies $A \leq^p_\beta B$. If the answer to this question is "true", we say that $\leq^p_\beta$ is stronger than $\leq^p_\alpha$. (Note that "stronger" does not in this paper necessarily promise "strictly stronger"—for example, each reduction is, under our definition, stronger than itself, but obviously is not strictly stronger than itself. Similarly, "weaker" does not necessarily promise "strictly weaker.") Roughly speaking, we show that if $\leq^p_\alpha$ and $\leq^p_\beta$ are chosen from among the Turing reductions, truth-table reductions, 2-truth-table reductions, and the set of query-monotonic reductions that we study, then the only "is stronger than" relationships that hold are the ones that obviously follow directly from the definitions. The rest provably do not hold. For example, a query-increasing reduction is by definition also a query-nondecreasing reduction (since each length-increasing query sequence is also a length-nondecreasing query sequence). Thus, query-nondecreasing reductions are stronger than query-increasing reductions. We prove that the converse is not true. That is, we prove that query-increasing reductions are strictly weaker than query-nondecreasing reductions.

---

[1] More broadly, the differences between various polynomial-time reducibilities have been extensively studied ever since the seminal work of Ladner, Lynch, and Selman [LLS75]. For an overview of this line of research, we refer the reader to the survey articles by Homer [Hom90,Hom97], Buhrman and Torenvliet [BT94], and Pavan [Pav03].

We now mention more specifically our results on the relative power of different reductions that we study in this paper. We prove that there are sets $A$ and $B$ such that $A \leq^p_{2\text{-}tt} B$, yet $A \not\leq^p_{li\text{-}T} B$ and $A \not\leq^p_{ld\text{-}T} B$. Since each 2-truth-table-reduction, query-increasing Turing reduction, and query-decreasing Turing reduction is also a Turing reduction, it certainly follows that in fact both $\leq^p_{li\text{-}T}$ and $\leq^p_{ld\text{-}T}$ are strictly weaker than $\leq^p_T$. However, we also prove that there are sets $C$ and $D$ such that $C \leq^p_{li\text{-}T} D$, $C \leq^p_{ld\text{-}T} D$, yet $C \not\leq^p_{tt} D$. Thus, each of $\leq^p_{2\text{-}tt}$ and $\leq^p_{tt}$ are, strength-wise, incomparable with each of $\leq^p_{li\text{-}T}$ and $\leq^p_{ld\text{-}T}$; neither in general guarantees that the other holds. In addition, we show that query-nonincreasing Turing reductions are strictly stronger than query-decreasing Turing reductions. Similarly, we prove that query-nondecreasing Turing reductions are strictly stronger than query-increasing Turing reductions. It is clear from the definitions that, for each $A, B \subseteq \Sigma^*$, $A \leq^p_{tt} B \implies A \leq^p_{lni\text{-}T} B$ and $A \leq^p_{lni\text{-}T} B \implies A \leq^p_T B$. However, note that from the results mentioned above (namely that each of $\leq^p_{li\text{-}T}$ and $\leq^p_{ld\text{-}T}$ are incomparable with $\leq^p_{tt}$ and that $\leq^p_{li\text{-}T}$ and $\leq^p_{ld\text{-}T}$ are incomparable) it follows that there are sets $B$ and $C$ such that $B \leq^p_{lni\text{-}T} C$ yet $B \not\leq^p_{tt} C$, there are sets $B$ and $C$ such that $B \leq^p_T C$ yet $B \not\leq^p_{lni\text{-}T} C$, there are sets $B$ and $C$ such that $B \leq^p_{lnd\text{-}T} C$ yet $B \not\leq^p_{tt} C$, and there are sets $B$ and $C$ such that $B \leq^p_T C$ yet $B \not\leq^p_{lnd\text{-}T} C$. That is, each of $\leq^p_{lni\text{-}T}$ and $\leq^p_{lnd\text{-}T}$ is strictly stronger than $\leq^p_{tt}$ and strictly weaker than $\leq^p_T$. Nonetheless, we prove that for each class $\mathcal{C}$ closed downward under polynomial-time many-one reductions (for example, NP, BPP, ⊕P, and PP), the reduction downward closures of $\mathcal{C}$ with respect to query-nonincreasing, query-nondecreasing, and Turing reductions are identical.

It is clear from the definitions that query-monotonic Turing reductions are no less restrictive than Turing reductions. Furthermore, from our results mentioned above it is clear that in some cases the query-increasing and query-decreasing restrictions have teeth. Thus, it is interesting to ask whether there are cases in which the restriction is toothless. In particular, we ask the following question: What structural properties should a set $S$ have so that each set that is Turing reducible to $S$ is in fact also query-increasing (query-decreasing) Turing reducible to $S$? We show that for a large class of sets—namely those that are paddable via a (polynomial-time) function whose output size is bounded tightly both from above and from below in the length of the input—the query-increasing (query-decreasing) restriction does not limit the power of polynomial-time oracle Turing machines using these sets as their database. We call these sets *tight paddable*. On one hand, we prove that there are NP-complete sets that are not tight paddable, but on the other hand hand, we show that many natural NP-complete problems (for example, satisfiability, vertex cover, and maximum clique) are tight paddable.

Glaßer et al. [GOP+05] have recently shown that all NP-complete sets are many-one autoreducible. A many-one autoreduction is a many-one reduction $\sigma$ from a set to itself such that, for each $x$, $\sigma(x) \neq x$. This is a very powerful result, but results on autoreducibility or length-increasing self-reducibility (a many-one reduction from a set $A$ to itself in which the function witnessing the reduction maps to strings that are strictly longer than the input string)[2] do not seem to give our result, though it is easy to have the impression that they do via a flawed argument. The flawed argument is this. Suppose one knows that each NP-complete set is many-one length-increasing self-reducible (as a side comment, we mention that this implies that P and NP differ). Given a set that $\leq^p_T$-reduces to some NP-complete set A, we try to convert that $\leq^p_T$-reduction to an query-increasing reduction by leaving the first query intact, and taking the second query and via the many-one length-increasing self-reduction mapping it to a string at most polynomially longer than the first string. And we map the third string that the original reduction would like to query to, via the many-one length-increasing self-reduction, a string at most polynomially longer than the string that we remapped the second query into. And so on. The problem with this approach is that the lengths snowball, and quickly become exponential. In contrast, "tight" padding, which we introduce in this paper, precisely avoids this type of snowballing explosion.

---

[2]Many-one length-decreasing self-reducibility when carelessly defined is possessed by no set, and when carefully defined—namely, in terms of the $\leq^p_{\widehat{m}}$ reduction of Ambos-Spies ([Amb87], see also [DGM94])—holds for precisely the sets in P and so is not interesting.

This paper is organized as follows. Sections 2 and 3 define the different query-monotonic reductions that are used in this paper, and show that all these reductions are robust in a certain sense. In Section 4, we compare the power of query-monotonic reductions with that of Turing and truth-table reductions. In Section 5, we study the query-monotonic reduction closures of sets in NP. In Section 6, we study tight paddability and prove that many important NP-complete sets are tight paddable.

## 2 Preliminaries

The alphabet for all strings, unless otherwise mentioned, is $\Sigma = \{0, 1\}$. For each $x \in \Sigma^*$, $rank(x)$ denotes the lexicographic rank of $x$, i.e., $rank(\epsilon) = 1$, $rank(0) = 2$, $rank(1) = 3$, $rank(00) = 4$, etc. The length of a string $x$ is denoted by $|x|$. For each string $x$, and for each $i$ such that $1 \leq i \leq |x|$, let $bit(x, i)$ denote the $i$th bit of $x$. For example, $bit(011, 1) = 0$. For each string $x$, and for each $i$ such that $0 \leq i \leq |x|$, let $pre(x, i)$ denote the prefix of $x$ of length $i$. For example, $pre(011, 0) = \epsilon$ and $pre(011, 2) = 01$.

$\langle \cdot, \cdot, \ldots, \cdot \rangle$ represents a one-to-one, polynomial-time computable and polynomial-time invertible multiarity pairing function (see [HHT97]) defined as follows. Let $\alpha(0) = 00$ and $\alpha(1) = 01$. Let $\beta(x) = \alpha(bit(x, 1))\alpha(bit(x, 2))\ldots\alpha(bit(x, |x|))$. Now define $\langle x_1, x_2, \ldots, x_k \rangle$ as $11\beta(x_1)11\beta(x_2)\ldots 11\beta(x_k)$. Note that $|\langle x_1, x_2, \ldots, x_k \rangle|$ is exactly $2(k + |x_1| + |x_2| + \ldots + |x_k|)$. Also note that, for each $x$, $10x$ has no preimage in $\langle \cdot, \cdot, \ldots, \cdot \rangle$. Due to the standard one-to-one mapping between strings and integers, we can apply the pairing function defined above to natural numbers, for example, by first converting each number to its binary representation and then applying the pairing function to these strings.

All sets, unless otherwise stated, are considered subsets of $\Sigma^*$. For each set $X$, $\|X\|$ denotes the cardinality of $X$. $\Sigma^n$ denotes the set of strings in $\Sigma^*$ of length exactly $n$. For any $C$ and $n$, $C^{=n} = \{a \mid a \in C \wedge |a| = n\}$, $C^{\leq n} = \{a \mid a \in C \wedge |a| \leq n\}$, $C^{<n} = \{a \mid a \in C \wedge |a| < n\}$, $C^{\geq n} = \{a \mid a \in C \wedge |a| \geq n\}$, and $C^{>n} = \{a \mid a \in C \wedge |a| > n\}$. For each $A \subseteq \Sigma^*$, $\chi_A$ is the characteristic function of $A$, i.e., for each $x \notin A$, $\chi_A(x) = 0$, and for each $x \in A$, $\chi_A(x) = 1$. By $\log n$ we mean $\log_2 n$.

We use the standard Turing machine model as described, for example, in [HU79]. We view all Turing machines as potentially taking an oracle, and so write "Turing machine" rather than "oracle Turing machine." Let $M_1, M_2, \ldots$ be a standard enumeration of deterministic polynomial-time Turing machines such that the running time of $M_i$ is bounded by $n^i + i$. Let $N_1, N_2, \ldots$ be a standard enumeration of nondeterministic polynomial-time Turing machines such that the running time of $N_i$ is bounded by $n^i + i$. Note that we build into our requirements of the standard enumerations the fact that Turing machines in the enumerations defined above run in time that is bounded by a polynomial that is *independent* of the oracle attached to the machine.[3] Thus, when we say that $M$

---

[3]This is certainly not true of all Turing machines. There has been some study of the difference between the power of machines that have a robust—that is, independent of the oracle—polynomial-time bound on their running time and those that run in polynomial-time for each oracle yet have no robust polynomial-time bound on their running

is a polynomial-time Turing machine, we mean that there is a $k \in \mathbb{N}$ such that, for each $X \subseteq \Sigma^*$, each $Y \subseteq \Sigma^*$, and for each $x \in \Sigma^*$, $M^{X,Y}(x)$ runs in time $|x|^k + k$ (thus so does $M^X(x)$ "=" $M^{X,\emptyset}(x)$ and $M(x)$ "=" $M^{\emptyset,\emptyset}(x)$). (For each Turing machine $M$ and for each $Y, Z \subseteq \Sigma^*$, $M^{Y,Z}$ represents machine $M$ with $Y$ as its first oracle and $Z$ as its second oracle. See [HHH97,HHH98,HHW99] for earlier uses of double oracles—there in the context of studying the effect of the order in which the two databases are accessed.)

For any Turing machine $N$ and any $x \in \Sigma^*$, we will use $N(x)$ as an abbreviation for "the computation of $N$ on $x$." We will use DPTM as an abbreviation for "deterministic polynomial-time Turing machine." We will use NPTM as an abbreviation for "nondeterministic polynomial-time Turing machine." Due to the properties of machines in our enumerations, a DPTM (NPTM) by definition runs in (polynomial) time that is independent of the oracle attached to it. For any NPTM $N$, $\#acc_N$ is the function such that, for any $x \in \Sigma^*$, $\#acc_N(x)$ is equal to the number of accepting computation paths of $N(x)$.

We now formally define polynomial-time truth-table reductions.

**Definition 2.1** *[LLS75] Let A and B be arbitrary sets.*

1. *We say that $A \leq_{tt}^p B$ (A polynomial-time truth-table reduces to B) if there exists a DPTM $M$ and a polynomial-time computable function $f$ such that, for each $x$, there exists an integer $m$ such that*

    (a) $f(x) = \langle q_1, q_2, \ldots, q_m \rangle$, and

    (b) $M(\langle x, \chi_B(q_1), \chi_B(q_2), \ldots, \chi_B(q_m) \rangle)$ accepts if and only if $x \in A$.

2. *For each $h : \mathbb{N} \to \mathbb{N}$, we say that $A \leq_{h(n)\text{-}tt}^p B$ (A polynomial-time $h(n)$-truth-table reduces to B) if there exists a DPTM $M$ and a polynomial-time computable function $f$ such that, for each $x$, there exists an integer $m \leq h(|x|)$ such that*

    (a) $f(x) = \langle q_1, q_2, \ldots q_m \rangle$, and

    (b) $M(\langle x, \chi_B(q_1), \chi_B(q_2), \ldots, \chi_B(q_m) \rangle)$ accepts if and only if $x \in A$.

As is standard, we will use $A \leq_{k\text{-}tt}^p B$ when we formally mean $A \leq_{(\lambda x.k)\text{-}tt}^p B$.

For each $a$ and $b$ such that $\leq_a^b$ is defined, and for each $A \subseteq \Sigma^*$, $\mathrm{R}_a^b(A)$ denotes $\{L \mid L \leq_a^b A\}$, the $\leq_a^b$-reduction downward closure of $A$. For each class $\mathcal{C}$ of languages and each $a$ and $b$ such that $\leq_a^b$ is defined, $\mathrm{R}_a^b(\mathcal{C}) = \bigcup_{A \in \mathcal{C}} \mathrm{R}_a^b(A)$. For any $a$ and $b$ such that $\leq_a^b$ is a defined reduction, and any class $\mathcal{C}$, we say that a set $B$ is $\mathcal{C}$-$\leq_a^b$-hard exactly if $\mathcal{C} \subseteq \mathrm{R}_a^b(B)$. For any $a$ and $b$ such that $\leq_a^b$ is a defined reduction, and any class $\mathcal{C}$, we say that a set $B$ is $\mathcal{C}$-$\leq_a^b$-complete if $B$ is $\mathcal{C}$-$\leq_a^b$-hard and $B \in \mathcal{C}$.

We now introduce query-monotonic reductions. We in fact define a general "query-restricted" Turing reduction, $\leq_{\rho\text{-}T}^p$, where $\rho$ is a restriction on the allowed sequences. Query-monotonic reductions are defined in terms of query-restricted Turing reductions: Each query-monotonic reduction is a query-restricted Turing reduction for some particular restriction $\rho$.

**Definition 2.2** *Let $\rho \subseteq \Sigma^* \cup \Sigma^* \times \Sigma^* \cup \Sigma^* \times \Sigma^* \times \Sigma^* \cup \cdots$.*

1. *For each Turing machine $M$, each $B \subseteq \Sigma^*$, and each $x \in \Sigma^*$, we say that $M$ has the $\rho$ query property with respect to $B$ on input $x$ if the sequence $q_1, q_2, \ldots, q_k$ of queries made by $M^B(x)$ to its oracle satisfies: $(x, q_1, q_2, \ldots, q_k) \in \rho$. (For example, $M^B(x)$ may legally ask no queries only if $(x) \in \rho$.)*

---

time [CHW99, Section 6].

2. For each Turing machine $M$ and each $B \subseteq \Sigma^*$, we say that $M$ has the $\rho$ query property with respect to $B$ if, for each $x \in \Sigma^*$, $M$ has the $\rho$ query property with respect to $B$ on input $x$.

3. For each $x \in \Sigma^*$, the set of valid query sequences in $\rho$ on input $x$ is the set $\{(x, q_1, q_2, \ldots, q_r) \,|\, (x, q_1, q_2, \ldots, q_r) \in \rho\}$.

4. $A \leq^p_{\rho\text{-}T} B$ if there exists a DPTM $M$ such that

   (a) $M$ has the $\rho$ query property with respect to $B$, and
   
   (b) $L(M^B) = A$.

We now define several query-monotonic restrictions that will be of interest to us. Note that the "$\Sigma^*$" part in the following definitions makes it legal for a Turing machine to ask no queries to its oracle.

**Definition 2.3**  1. (Length-increasing) $\rho_{li} = \Sigma^* \cup \{(x, q_1, q_2, \ldots, q_k) \,|\, k \geq 1 \wedge |q_1| < |q_2| < \ldots < |q_k|\}$.

2. (Length-decreasing) $\rho_{ld} = \Sigma^* \cup \{(x, q_1, q_2, \ldots, q_k) \,|\, k \geq 1 \wedge |q_1| > |q_2| > \ldots > |q_k|\}$.

3. (Length-nonincreasing) $\rho_{lni} = \Sigma^* \cup \{(x, q_1, q_2, \ldots, q_k) \,|\, k \geq 1 \wedge |q_1| \geq |q_2| \geq \ldots \geq |q_k|\}$.

4. (Length-nondecreasing) $\rho_{lnd} = \Sigma^* \cup \{(x, q_1, q_2, \ldots, q_k) \,|\, k \geq 1 \wedge |q_1| \leq |q_2| \leq \ldots \leq |q_k|\}$.

5. (Strong length-increasing) $\rho_{s\text{-}li} = \Sigma^* \cup \{(x, q_1, q_2, \ldots, q_k) \,|\, k \geq 1 \wedge |x| < |q_1| < |q_2| < \ldots < |q_k|\}$.

6. (Strong length-decreasing) $\rho_{s\text{-}ld} = \Sigma^* \cup \{(x, q_1, q_2, \ldots, q_k) \,|\, k \geq 1 \wedge |x| > |q_1| > |q_2| > \ldots > |q_k|\}$.

7. (Strong length-nonincreasing) $\rho_{s\text{-}lni} = \Sigma^* \cup \{(x, q_1, q_2, \ldots, q_k) \,|\, k \geq 1 \wedge |x| \geq |q_1| \geq |q_2| \geq \ldots \geq |q_k|\}$.

8. (Strong length-nondecreasing) $\rho_{s\text{-}lnd} = \Sigma^* \cup \{(x, q_1, q_2, \ldots, q_k) \,|\, k \geq 1 \wedge |x| \leq |q_1| \leq |q_2| \leq \ldots \leq |q_k|\}$.

For each $\alpha \in \{li, ld, lnd, lni, s\text{-}li, s\text{-}ld, s\text{-}lni, s\text{-}lnd\}$, and for each $A, B \subseteq \Sigma^*$, we will abuse notation and use $A \leq^p_{\alpha\text{-}T} B$ when we formally mean $A \leq^p_{\rho_\alpha\text{-}T} B$. For each Turing machine $M$ and each $B \subseteq \Sigma^*$, if $M$ has the $\rho_{li}$ (respectively, $\rho_{ld}$, $\rho_{lni}$, $\rho_{lnd}$, $\rho_{s\text{-}li}$, $\rho_{s\text{-}ld}$, $\rho_{s\text{-}lni}$, $\rho_{s\text{-}lnd}$) query property with respect to $B$, then we say that $M$ has the query-increasing (respectively, query-decreasing, query-nonincreasing, query-nondecreasing, strong query-increasing, strong query-decreasing, strong query-nonincreasing, strong query-nondecreasing) property with respect to $B$. If $A \leq^p_{li\text{-}T} B$ (respectively, $A \leq^p_{ld\text{-}T} B$, $A \leq^p_{lni\text{-}T} B$, $A \leq^p_{lnd\text{-}T} B$, $A \leq^p_{s\text{-}li\text{-}T} B$, $A \leq^p_{s\text{-}ld\text{-}T} B$, $A \leq^p_{s\text{-}lni\text{-}T} B$, $A \leq^p_{s\text{-}lnd\text{-}T} B$), then we say that $A$ polynomial-time query-increasing (respectively, query-decreasing, query-nonincreasing, query-nondecreasing, strong query-increasing, strong query-decreasing, strong query-nonincreasing, strong query-nondecreasing) Turing reduces to $B$. Sometimes, for succinctness and to avoid repetition, we will not explicitly include the word "Turing," e.g., we will write "query-nonincreasing reduction" for "query-nonincreasing Turing reduction."

We now define relativized query-monotonic reductions. For each $\rho \subseteq \Sigma^* \cup \Sigma^* \times \Sigma^* \cup \Sigma^* \times \Sigma^* \times \Sigma^* \cup \cdots$, we say that, relative to $Z$, $M$ has the $\rho$ query property with respect to $Y$ if, for each $x \in \Sigma^*$, the sequence $q_1, q_2, \ldots, q_k$ of queries that $M^{Y,Z}(x)$ asks to its first oracle satisfies $(x, q_1, q_2, \ldots, q_k) \in \rho$. $A \leq^{p,C}_{\rho\text{-}T} B$ if there exists a DPTM $M$ such that $L(M^{B,C}) = A$ and, relative to $C$, $M$ has the $\rho$ query property with respect to $B$.

Though it is slightly broken English, for succinctness we will at times write phrases using the locution "queries [string] to [oracle being queried]." For example, we will at times write "the machine then queries $x$ to $A$" for "the machine then asks the query $x$ to the oracle $A$."

We now define a new notion of padding—*tight paddability*—that we will use to understand query-monotonic reductions over NP. Before we define tight paddability, for comparison we review notions of padding that are standard in the literature. Paddability of sets has been used in complexity theory in many contexts including the seminal work of Berman and Hartmanis [BH77] on polynomial-time isomorphism for sets in NP, and of Hartmanis [Har78] on the study of logspace isomorphism for sets in NL, CSL, P, NP, PSPACE, etc. Mahaney and Young [MY85] use padding functions to prove results regarding the structure of polynomial-time many-one degrees.

A (polynomial-time) paddable set is one for which there is a polynomial-time function (called a padding function for that set) that allows one to map the input string to a longer string while preserving membership in the set.

**Definition 2.4** *Let $A \subseteq \Sigma^*$. Then $\sigma : \Sigma^* \to \Sigma^*$ is a* padding function *for $A$ if*

1. *$\sigma$ is polynomial-time computable,*

2. *for each $x \in \Sigma^*$, $\sigma(x) \in A$ if and only if $x \in A$, and*

3. *for each $x \in \Sigma^*$, $|\sigma(x)| > |x|$.*

*We say that $A$ is* paddable *if there exists a padding function for $A$.*

Berman and Hartmanis [BH77] define two different types of functions called Z-padding functions and S-padding functions (though, formally speaking S-padding functions need not be padding functions in the sense of Definition 2.4).

**Definition 2.5** *[BH77] Let $A \subseteq \Sigma^*$. Then $\sigma : \Sigma^* \to \Sigma^*$ is a* Z-padding function *for $A$ if $\sigma$ is a padding function and $\sigma$ is a one-to-one function. We say that $A$ is* Z-paddable *if there exists a Z-padding function for $A$.*

**Definition 2.6** *[BH77] Let $A \subseteq \Sigma^*$. Then $\sigma : \Sigma^* \times \Sigma^* \to \Sigma^*$ is an* S-padding function *for $A$ if*

1. *$\sigma$ is polynomial-time computable,*

2. *$\sigma$ is polynomial-time invertible in the second argument, i.e., there is a polynomial-time computable function $\sigma'$ such that, for each $x, y \in \Sigma^*$, $\sigma'(\sigma(x, y)) = y$, and*

3. *for each $x, y \in \Sigma^*$, $\sigma(x, y) \in A$ if and only if $x \in A$.*

*We say that $A$ is* S-paddable *if there exists an S-padding function for $A$.*

Mahaney and Young [MY85] note that any set that is S-paddable is S-paddable via some function that is polynomial-time invertible in both arguments, i.e., there exists a polynomial-time computable function $\rho$ such that for each $x$ and $y$, $\rho(\sigma(x, y)) = \langle x, y \rangle$. Also, note that if a function is invertible in both arguments, in the particular sense just mentioned, then it is one-to-one. We state these properties in Proposition 2.7.

**Proposition 2.7** *(essentially [MY85]) Let $A$ be S-paddable. Then there is an S-padding function $\pi$ for $A$ such that*

1. *$\pi$ is polynomial-time invertible in both arguments (in the sense defined above),*

2. *$\pi$ is honest (that is, there is a polynomial $q$ such that, for each $x, y \in \Sigma^*$, $|x|+|y| \leq q(|\pi(x,y)|)$.*

3. *$\pi$ is one-to-one, that is, for each $a$, $b$, $c$, and $d$, $\pi(a,b) = \pi(c,d)$ only if $a = c$ and $b = d$,*

4. for each $x, y \in \Sigma^*$, $|\pi(x, y)| > |x|$.

**Proof** Let $\sigma$ be a S-padding function for $A$. By definition, $\sigma$ is polynomial-time invertible in the second argument. Let $\sigma'$ be the function that inverts $\sigma$ in the second argument. We will first define a function $\alpha : \Sigma^* \times \Sigma^* \to \Sigma^*$ that satisfies the first three conditions and then use $\alpha$ to define a function $\pi : \Sigma^* \times \Sigma^* \to \Sigma^*$ that satisfies all the four conditions.

Consider $\alpha$ defined as follows: $\alpha(x, y) = \sigma(x, \langle x, y \rangle)$. $\alpha$ is polynomial-time computable because $\sigma$ is polynomial-time computable. $\alpha$ is polynomial-time invertible in both arguments (in the sense mentioned above) via $\sigma'$ because, for each $x, y \in \Sigma^*$, $\sigma'(\alpha(x, y)) = \sigma'(\sigma(x, \langle x, y \rangle)) = \langle x, y \rangle$, where the last equality follows from the definition of $\sigma'$. Thus, $\alpha$ is a polynomial-time invertible-in-both-arguments (in the sense defined above) S-padding function for $A$. Like all functions invertible in both arguments (in the above sense) functions, it is one-to-one.

Since $\alpha$ is invertible in both arguments, there is a polynomial $q'$ such that, for each $x$ and $y$, $|\langle x, y \rangle| = |\alpha^{-1}(\alpha(x, y))| \leq q'(|\alpha(x, y)|)$. Thus, there is a strictly increasing polynomial $q$ such that $|x| + |y| \leq q(|\alpha(x, y)|)$. Equivalently, $|\alpha(x, y)| \geq q^{-1}(|x| + |y|)$. In other words, $\alpha$ is honest, i.e., it does not *shrink* its input by more than an inverse-polynomial factor.

For each $x, y \in \Sigma^*$, define $\pi(x, y) = \alpha(x, \langle y, 0^{q(|x|)+1} \rangle)$. Clearly, for each $x \in \Sigma^*$, $|\pi(x, y)| > |x|$. Also, $\pi$ is an S-padding function for $A$ because we can show, using the fact that $\alpha$ is an S-padding function for $A$, that $\pi$ obeys all the three conditions mentioned in Definition 2.6. It is easy to see that since $\alpha$ is honest, one-to-one, and polynomial-time invertible in both arguments, so is $\pi$. ❑

It is important to note that any S-paddable set $A$ is also Z-paddable. Note that the $\pi$'s created in Proposition 2.7 are honest, that is, they can "shrink" the size of their inputs at most polynomially. We can use this property to construct a Z-padding function for $A$. In particular, $\rho(x) = \pi(x, 0^{q(|x|)-|x|})$ is a Z-padding function for $A$, where $\pi$ is an invertible-in-both-arguments S-padding function for the set (e.g., via Proposition 2.7) and $q$ is the associated honesty polynomial. Note that most natural NP-complete problems (for example, SAT, vertex cover, max clique, 3-colorability, etc.) are obviously S-paddable.

We now define two notions of tight padding: *tight paddability* and *tight Z-paddability*. Informally speaking, a tight padding function for a set is a padding function that has strong guarantees on the length of its output.

**Definition 2.8** *Let $A \subseteq \Sigma^*$.*

1. *Let $\sigma : \Sigma^* \to \Sigma^*$. Then $\sigma$ is a tight padding function for $A$ if $\sigma$ is a padding function for $A$ and there exists a $k \in \mathbb{N}$ such that, for each $x \in \Sigma^*$, $|\sigma(x)| \leq |x| + k$. We say that $A$ is tight paddable if there is a tight padding function for $A$.*

2. *Let $\sigma : \Sigma^* \to \Sigma^*$. Then $\sigma$ is a tight Z-padding function for $A$ if $\sigma$ is a Z-padding function for $A$ and there exists a $k \in \mathbb{N}$ such that, for each $x \in \Sigma^*$, $|\sigma(x)| \leq |x| + k$. We say that $A$ is tight Z-paddable if there is a tight Z-padding function for $A$.*

Clearly, each tight paddable set is also paddable. Similarly, each tight Z-paddable set is also Z-paddable. Figure 1 shows the relationships among the different notions of paddability.

One might ask why we have not defined a "tight" analog of the S-paddable sets. One could. But we mention in passing that no one-to-one 2-ary function satisfies the length restriction associated with tightness. Let $\sigma$ be an arbitrary one-to-one 2-ary function. Let $n \in \mathbb{N}$. The number of pairs of strings $x$ and $y$ such that $|x|+|y|$ is $n$ is exactly $(n+1)2^n$. Thus, the length of the longest string whose preimage $(x, y)$ in $\sigma$ is such that $|x|+|y| = n$ is at least $\lceil \log(1+(n+1)2^n) - 1 \rceil \geq n-1+\log(n+1)$. Thus, for any such $\sigma$ and for each $k \in \mathbb{N}$, there exist $x$ and $y$ such that $\sigma(x, y) > |x| + |y| + k$.

In Section 6, we show that many well-known NP-complete problems such as satisfiability of boolean formulas, vertex cover in graphs, and size of the largest clique in graphs are tight Z-paddable.

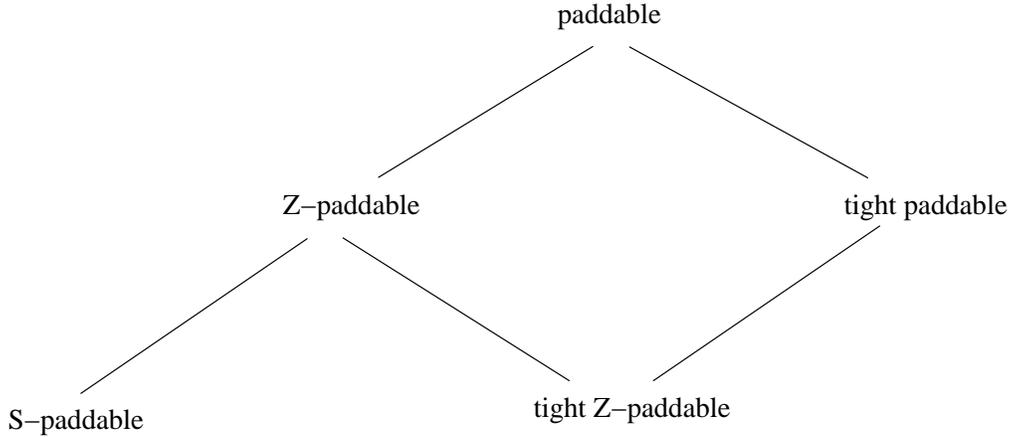

Figure 1: Relationships between different notions of paddability. A line between $a$ and $b$ such that $a$ lies above $b$ indicates that any set of type $b$ is also a set of type $a$.

We also show that many types of query-monotonic reductions are equivalent (and in fact, also equivalent to Turing reductions) when the set being reduced to is tight paddable.

## 3 Robustness of Query-Monotonic Reductions

Note that in the definition of $\leq^p_{\rho\text{-}T}$ reductions (Definition 2.2, part 4), the machine witnessing the reduction need only have the $\rho$ query property when its database (oracle) is the particular set being reduced to. For example, a machine $M$ witnessing a $\leq^p_{li\text{-}T}$ reduction from $A$ to $B$ is allowed to disobey the $\rho_{li}$ query property when the oracle that the machine accesses is not $B$. That is, for some $B' \neq B$, $M$ might not have the query-increasing property with respect to $B'$. Now, consider a potentially more restrictive form of query-monotonic reductions in which the machine witnessing the reduction is required to be query-monotonic *robustly*, i.e., not just for the set it is witnessing the reduction to, but also for *every* set. We formalize this notion in Definition 3.1 below. We will soon show that whether such robustness does limit the power of query-restricted Turing reductions is dependent both on the constraint $\rho$ and on whether P = NP.

**Definition 3.1** *Let $\rho \subseteq \Sigma^* \cup \Sigma^* \times \Sigma^* \cup \Sigma^* \times \Sigma^* \times \Sigma^* \cup \cdots$.*

1. *For each Turing machine $M$, we say that $M$ has the robust $\rho$ query property if, for each $C \subseteq \Sigma^*$, $M$ has the $\rho$ query property with respect to $C$.*

2. *For each $A, B \subseteq \Sigma^*$, $A \leq^p_{r\text{-}\rho\text{-}T} B$ if there exists a DPTM $M$ such that*

    (a) *$M$ has the robust $\rho$ query property, and*
    (b) *$L(M^B) = A$.*

Informally, this is simply part 4 of Definition 2.2 except with the $\rho$ query property being global. For each $\alpha \in \{li, ld, lni, lnd, s\text{-}li, s\text{-}ld\ s\text{-}lni, s\text{-}lnd\}$, and for each $A, B \subseteq \Sigma^*$, we will abuse notation and use $A \leq^p_{r\text{-}\alpha\text{-}T} B$ when we formally mean $A \leq^p_{r\text{-}\rho_\alpha\text{-}T} B$. If $A \leq^p_{r\text{-}li\text{-}T} B$ (respectively $A \leq^p_{r\text{-}ld\text{-}T} B$, $A \leq^p_{r\text{-}lni\text{-}T} B$, $A \leq^p_{r\text{-}lnd\text{-}T} B$, $A \leq^p_{r\text{-}s\text{-}li\text{-}T} B$, $A \leq^p_{r\text{-}s\text{-}ld\text{-}T} B$, $A \leq^p_{r\text{-}s\text{-}lni\text{-}T} B$, $A \leq^p_{r\text{-}s\text{-}lnd\text{-}T} B$), then we say that *A robust query-increasing* (respectively, *robust query-decreasing*, *robust query-nonincreasing*, *robust query-nondecreasing*, *robust strong query-increasing*, *robust strong query-decreasing*, *robust strong query-nonincreasing*, *robust strong query-nondecreasing*) *Turing reduces to B*.

Thus, robust query-monotonic (for example, query-increasing or query-decreasing) Turing machines have the query-monotonic property with respect to all oracles and all input strings, while query-monotonic Turing machines are only required to have the query-monotonic property with respect to one oracle and all input strings. The issue of whether Turing machines have properties robustly (for all oracles) has been studied in literature, for example, in the the study of positive Turing reductions [Sel79,HJ91], in the study of robust Turing machines and helper sets [Sch85, Ko87,CHV93], and even regarding whether a Turing machine runs in polynomial time [CHV93]. Regarding the first of these, various forms of positive truth-table reductions and positive Turing reductions (neither of which we define here) are used, among other things, to study the P-selective sets (see [Sel79,HJ91]). It is now known that positive and robustly positive reductions are different under standard complexity-theoretic assumptions and under suitable relativizations. In contrast, we prove via Theorem 3.2 that robust query-monotonic Turing reductions are no more restrictive than query-monotonic Turing reductions. Informally speaking, Theorem 3.2 says that reductions based on "nice" polynomial-time-recognizable query-constraint collections are just as powerful when restricted to robustly obeying the constraints, i.e., obeying them not just for the set being reduced to, but with respect to every set. Our niceness constraint here—which is quite natural and is satisfied by all the query-monotonic reduction types we have defined—is that all the nonempty prefixes of any nonempty, legal query-vector are themselves legal. This disallows a potentially fatal case—the case where the oracle itself leads the computation through a minefield of illegal vectors, but at the end reaches a legal one. We will, after Corollary 3.3, further discuss the importance (and potential nonimportance) of the niceness condition, and in particular will show that on one hand if P = NP then for P-recognizable query properties robustness can be achieved always (even for $\rho$ that violate niceness), but that on the other hand there are (necessarily non-nice, non-P-recognizable) query properties $\rho$ such that in appropriate relativized worlds, $\rho$ and its robust counterpart differ.

**Theorem 3.2** *Let $\rho \subseteq \Sigma^* \cup \Sigma^* \times \Sigma^* \cup \Sigma^* \times \Sigma^* \times \Sigma^* \cup \cdots$ be a set of tuples such that $\rho' = \{\langle x_1, x_2, \ldots, x_k \rangle \mid (x_1, x_2, \ldots, x_k) \in \rho\} \in \mathrm{P}$ and $\rho$ satisfies the following: For each $k \geq 0$, and each $x, q_1, q_2, \ldots, q_{k+1} \in \Sigma^*$, if $(x, q_1, x_2, \ldots, x_{k+1}) \in \rho$, then $(x_1, x_2, \ldots, x_k) \in \rho$. Then, for each $A, B \subseteq \Sigma^*$, $A \leq^p_{\rho\text{-}T} B$ if and only if $A \leq^p_{r\text{-}\rho\text{-}T} B$.*

**Proof** The ($\Leftarrow$) direction is trivially true. We will show that the ($\Rightarrow$) direction also holds. Let $M$ be a DPTM witnessing $A \leq^p_{\rho\text{-}T} B$. We will show that $A \leq^p_{r\text{-}\rho\text{-}T} B$. Consider a Turing machine $M_1$ that, on input $x$, simulates $M(x)$ exactly, except when $M(x)$ is about to ask a query (say $q$) to its oracle. When $M(x)$ is about to ask a query to its oracle $M_1$ does the following. It computes all the queries that $M_1(x)$ has previously asked to its oracle in the current run. (Note that this can easily be computed, for example, by having $M_1$ keep track of the queries it asks of its oracle.) Let these queries be $q_1, q_2, \ldots, q_k$. $M_1(x)$ computes whether $z = (x, q_1, q_2, \ldots, q_k, q)$ is a member of $\rho$ by running the polynomial-time algorithm for $\rho'$ on the string $\langle x, q_1, q_2, \ldots, q_k, q \rangle$. (By hypothesis, such an algorithm exists.) If $z \notin \rho$, $M_1(x)$ halts and rejects. (Note that in this case the oracle attached to $M_1$ is not $B$ since, by hypothesis, all query prefixes of all sequences of queries asked by $M^B$ on all inputs are from the set $\rho$.) If $z \in \rho$, $M_1(x)$ asks query $q$ to its oracle and proceeds with the current computation.

Since $M$ is a DPTM, so is $M_1$. Also, by construction, $M_1$ has the robust $\rho$ query property, because $M_1$ only asks a query $q$ if the sequence of queries up to and including $q$ is a member of $\rho$. Since, by hypothesis, $M$ witnesses $A \leq^p_{\rho\text{-}T} B$, we have that, for all strings $x$, the query sequence of $M^B(x)$ is in $\rho$. Thus, by construction, for all $x$, $M_1^B(x)$ simulates $M^B(x)$ appropriately. So, $L(M_1^B) = L(M^B) = A$. Since $M_1$ has the robust $\rho$ query property, $M_1$ runs in polynomial time, and $L(M_1^B) = A$, it follows that $A \leq^p_{r\text{-}\rho\text{-}T} B$ (via $M_1$). ❑

Since $\rho_{li}$, $\rho_{ld}$, $\rho_{lni}$, $\rho_{lnd}$, $\rho_{s\text{-}li}$, $\rho_{s\text{-}ld}$, $\rho_{s\text{-}lni}$, and $\rho_{s\text{-}lnd}$ all satisfy the niceness condition, as an immediate corollary we have that for each query-monotonic Turing reduction in Definition 2.3 robustness is free.

**Corollary 3.3** *For each $\alpha \in \{li, ld, lni, lnd, s\text{-}li, s\text{-}ld, s\text{-}lni, s\text{-}lnd\}$, and for each $A, B \subseteq \Sigma^*$, $A \leq^p_{\alpha\text{-}T} B$ if and only if $A \leq^p_{r\text{-}\alpha\text{-}T} B$.*

Is the niceness condition—that nonempty prefixes of nonempty legal sequences must themselves be legal—needed for the robust $\rho$ query property to hold? The following two results show that if P = NP the condition is superfluous (in proving this our strategy will be to ensure that we always have an "escape route" available to us—as a set of additional queries that along with those we asked are an allowed query vector), and that there are oracle worlds where the condition is needed.

Before we state Theorem 3.4, we state a definition that will be used in its proof. Given $\rho \subseteq \Sigma^* \cup \Sigma^* \times \Sigma^* \cup \Sigma^* \times \Sigma^* \times \Sigma^* \cup \cdots$, sequences of strings $s = (x_1, x_2, \ldots, x_k)$ and $s' = (y_1, y_2, \ldots, y_t)$, and integer $r$, we say that $s'$ is a $(\rho, s, r)$ *escape route* if

1. $t \leq r$,

2. for each $1 \leq i \leq t$, $|y_t| \leq r$, and

3. $(x_1, x_2, \ldots, x_k, y_1, y_2, \ldots, y_t) \in \rho$.

The following property of escape routes follows from the definition above: Let $M$ be a Turing machine such that there is a polynomial $p$ bounding the running time of $M$ (regardless of the oracle). Assume that $M$ has the $\rho$ query property with respect to $B$. Let $x \in \Sigma^*$ and $x_1, x_2, \ldots, x_k$ be a prefix of queries that $M^B(x)$ asks to its oracle. Let $s = (x, x_1, x_2, \ldots, x_k)$. Then there must exist a $(\rho, s, p(|x|))$ escape route. We can now state Theorem 3.4 and its proof.

**Theorem 3.4** *Let $\rho \subseteq \Sigma^* \cup \Sigma^* \times \Sigma^* \cup \Sigma^* \times \Sigma^* \times \Sigma^* \cup \cdots$ be a set of tuples such that the set $\rho' = \{\langle x_1, x_2, \ldots, x_k \rangle \,|\, (x_1, x_2, \ldots, x_k) \in \rho\}$ is in P. If P = NP, then for each $A, B \subseteq \Sigma^*$, $A \leq^p_{\rho\text{-}T} B$ if and only if $A \leq^p_{r\text{-}\rho\text{-}T} B$.*

**Proof** The ($\Leftarrow$) direction is trivially true. We will show that the ($\Rightarrow$) direction also holds. Let $M$ be a DPTM witnessing $A \leq^p_{\rho\text{-}T} B$. Note that by definition (see Section 2), the running time of $M$ is bounded by a polynomial (say, $p$) independent of the oracle. We will show that $A \leq^p_{r\text{-}\rho\text{-}T} B$ by describing a Turing machine $M_1$ that witnesses $A \leq^p_{r\text{-}\rho\text{-}T} B$. $M_1$ will have the property that, for each oracle $C$, for each string $x$, and at each step in the computation $M_1^C(x)$, if $q_1, q_2, \ldots, q_k$ is the sequence of queries that have already been asked, then there exists a $(\rho, s, p(|x|))$ escape route, where $s = (x, q_1, q_2, \ldots, q_k)$. Note that since for each string $x$ there exists a $(\rho, (x), p(|x|))$ escape route (namely, the sequence of strings that $M^B(x)$ asks to its oracle), this property holds trivially at the very start of computation $M_1^C(x)$. We now describe $M_1$.

$M_1$, on input $x$, simulates $M(x)$, except in the following two cases: (a) when $M(x)$ is about to ask a query to its oracle, and (b) when $M(x)$ is about to halt. We will describe what $M_1$ does in each of the above two cases.

Consider case (a): $M(x)$ is about to ask a query to its oracle. In this case $M_1$ does the following. It computes all the queries that $M_1(x)$ has previously asked to its oracle. (Note that this can easily be computed, for example, by having $M_1$ keep track of the queries it asks to its oracle.) Let these queries be $q_1, q_2, \ldots, q_k$. Let $s_1 = (x, q_1, q_2, \ldots, q_k, q)$ and let $s_2 = (x, q_1, q_2, \ldots, q_k)$. $M_1(x)$ computes whether a $(\rho, s_1, p(|x|))$ escape route exists. (Clearly, computing the existence of a $(\rho, s_1, p(|x|))$ escape route can be done in NP. Since P = NP, by assumption, this can even be done deterministically in polynomial time.) If a $(\rho, s_1, p(|x|))$ escape route exists, then $M_1(x)$ asks query $q$ to its oracle and proceeds with the current computation. If no such escape route exists, it follows that the oracle attached to $M_1$ is not $B$ since, by hypothesis, $M$ has the $\rho$ query property with respect to $B$. In this case $M_1$ computes a $(\rho, s_2, p(|x|))$ escape route $e$. (Note that since $M_1$ queried $q_k$ at an earlier step, the construction ensures the existence of a $(\rho, s_2, p(|x|))$ escape route. Furthermore, since P = NP, this sequence can be computed deterministically in polynomial time.) Let $e = (y_1, y_2, \ldots, y_m)$. $M_1$ asks the following queries in sequence: $y_1, y_2, \ldots, y_m$. $M_1$ then halts and rejects.

Now consider case (b): $M(x)$ is about to halt. Let $q_1, q_2, \ldots, q_k$ be the sequence of queries that $M_1$ has asked to its oracle. Let $s = (x, q_1, q_2, \ldots, q_k)$. $M_1$ computes a $(\rho, s, p(|x|))$ escape route $e$. Clearly, such an escape route exists, since it exists at the start of the computation and, as argued in case (a) above, at each subsequent step. Let $e = (y_1, y_2, \ldots, y_m)$. (Note: When our oracle happens to be $B$, a legal escape sequence will always be to take $e$ to be an empty sequence.) $M_1$ asks the following queries in sequence: $y_1, y_2, \ldots, y_m$. If $M(x)$ was about to halt and accept, $M_1(x)$ halts and accepts. On the other hand, if $M(x)$ was about to halt and reject, $M_1(x)$ halts and rejects. This completes the description of $M_1$.

Since $M$ is a DPTM, so is $M_1$. Also, by construction, $M_1$ has an escape route available at each step, and just before halting (both in case (a) and in case (b)) it "takes" the escape route and ensures that the sequence of queries is in fact a valid one. Thus, $M_1$ has the robust $\rho$ query property. Also, it follows from construction that $L(M_1^B) = L(M^B) = A$. Since, $M_1$ has the robust $\rho$ query property, $M_1$ runs in polynomial time, and $L(M_1^B) = A$, it follows that $A \leq^p_{r\text{-}\rho\text{-}T} B$ (via $M_1$). ❑

Theorem 3.4 shows that for all easily computable query sequence collections, if P = NP then the niceness condition is not required for robustness. Thus, proving that the niceness condition is required for robustness will imply P $\neq$ NP, and thus cannot be proven using relativizable techniques. However, this leaves open the possibility that the niceness condition is *never* required. In Theorem 3.5 we show that proving that the niceness condition is never required would require nonrelativizable techniques, since there is an oracle $C$ and a constraint set $\rho \in \mathrm{P}^C$ such that $\leq^{p,C}_{\rho\text{-}T}$ and $\leq^{p,C}_{r\text{-}\rho\text{-}T}$ differ. Of course, in light of (the natural relativized version of) Theorem 3.2, the $\rho$ of Theorem 3.5 must be non-nice. In fact, Theorem 3.5 shows that the difference between $\leq^{p,C}_{\rho\text{-}T}$ and $\leq^{p,C}_{r\text{-}\rho\text{-}T}$ is witnessed by a simple set, namely the empty set. The reader might wonder why a Turing machine that rejects on all inputs (without asking any queries to its oracle) does not establish that, for each set $B$, $\emptyset \leq^{p,C}_{r\text{-}\rho\text{-}T} B$. The reason is that on some inputs asking no queries at all might not be legal in $\rho$.

**Theorem 3.5** *There is an oracle $C$, a set $B$, and a constraint set $\rho$ such that the set $\rho' = \{\langle x_1, x_2, \ldots, x_k \rangle \mid (x_1, x_2, \ldots, x_k) \in \rho\} \in \mathrm{P}^C$, $\emptyset \leq^{p,C}_{\rho\text{-}T} B$, and yet $\emptyset \not\leq^{p,C}_{r\text{-}\rho\text{-}T} B$.*

**Proof** We will construct $B$, $C$, and $\rho$ in stages. In Stage $i$, we will diagonalize against machine $M_i$. That is, in Stage $i$, we will show that $M_i$ does not witness $\emptyset \leq^{p,C}_{r\text{-}\rho\text{-}T} B$. Define

$$Z = \{2^{2^2}, 2^{2^{2^{2^2}}}, 2^{2^{2^{2^{2^{2^2}}}}}, \ldots\}.$$

We mention the following property of $Z$, which will be used below:

**Lemma 3.6** *Let $n$ and $n'$ be the $k$th and $(k+1)$st smallest numbers in $Z$. Then $n' > 2^{n+1} > 1 + n^k + k$.*

**Proof of Lemma 3.6** Immediate from the definition of $Z$. ❏ (Lemma 3.6)

We continue with the proof of Theorem 3.5. Define $D = \{0^n \mid n \in Z\}$. Let $B_0 = \emptyset$, $C_0 = \emptyset$, $\rho_0 = \{(x) \mid x \notin D\}$, and $n_0 = 1$.

**Stage $k \in \mathbb{N}^+$:** Let $M = M_k$. Let $n = n_k$ be the integer $2^{2^{2^{n_{k-1}}}}$. Let $B_{k-1}$, $C_{k-1}$, and $\rho_{k-1}$ be the set of strings added to $B$, $C$, and $\rho$, respectively, before the $k$th stage.

For each $i$ such that $1 \leq i \leq 2^n$, let $x_i$ be the lexicographically $i$th string in $\Sigma^n$. Let $P = \{x_1, x_2, \ldots, x_{n+1}\}$. For each $Y \subseteq \Sigma^n$, let $c_Y$ be the string $\chi_Y(x_1)\chi_Y(x_2)\ldots\chi_Y(x_{n+1})$. For each $Y \subseteq \Sigma^n$, let $q_Y$ be the string $\langle x_1, x_2, \ldots, x_{n+1}, c_Y\rangle$. That is, $q_Y$ represents a sequence of $n+2$ oracle queries where the first $n+1$ strings queried are the lexicographically first $n+1$ strings in $\Sigma^n$ and the last string is the $(n+1)$-bit string formed by concatenating the answers (obtained from $Y$) to the previous queries.

There are the following two cases:

1. Consider the case that there exist $Y, Y' \subseteq P$ such that the sequence of queries that $M^{B_{k-1} \cup Y', C_{k-1} \cup \{c_Y\}}(0^n)$ asks to its first oracle is not $q_Y$. In this case, set $B_k = B_{k-1} \cup Y$, set $C_k = C_{k-1} \cup \{q_Y\}$, and set $\rho_k = \rho_{k-1} \cup \{(0^n, x_1, x_2, \ldots, x_{n+1}, c_Y)\}$. Go to Stage $k+1$.

2. The remaining case is: For each $Y, Y' \subseteq P$, the sequence of queries that $M^{B_{k-1} \cup Y', C_{k-1} \cup \{c_Y\}}$ asks to its first oracle is $q_Y$. We will derive a contradiction. Let $Y \subseteq P$ be the lexicographically smallest set such that $M^{B_{k-1}, C_{k-1}}(0^n)$ does not ask $c_Y$ to its second oracle and the sequence of queries that $M^{B_{k-1}, C_{k-1}}(0^n)$ asks to its first oracle is not $q_Y$. Note that $Y$ is well defined because $M^{B_{k-1}, C_{k-1}}(0^n)$ asks at most $n^k + k$ queries, the number of distinct subsets of $P$ is $2^{n+1}$, and by Lemma 3.6 we have $2^{n+1} > 1 + n^k + k$. Since $M^{B_{k-1}, C_{k-1}}(0^n)$ does not ask $c_Y$ to its second oracle, the sequence of queries asked by $M^{B_{k-1}, C_{k-1} \cup \{c_Y\}}(0^n)$ is exactly the same as the sequence of queries asked by $M^{B_{k-1}, C_{k-1}}(0^n)$. By our choice of $Y$, we have that the sequence of queries asked by $M^{B_{k-1}, C_{k-1} \cup \{c_Y\}}(0^n)$ to its first oracle is not $q_Y$, which is a contradiction. Thus, this case cannot arise.

**End of Stage $k$.**

We now define $B = \bigcup_{k \geq 0} B_k$, $C = \bigcup_{k \geq 0} C_k$, and $\rho = \bigcup_{k \geq 0} \rho_k$. For each $x \in \Sigma^*$, there is exactly one valid query sequence in $\rho$ on input $x$. If $x \notin D$, $(x)$ is the only valid query sequence in $\rho$ on input $x$. If $x \in D$, then for each $q_1, q_2, \ldots, q_r \in \Sigma^*$, $(x, q_1, q_2, \ldots, q_r) \in \rho$ if and only if $r = |x| + 2$, for each $i < r$, $q_i$ is the lexicographically $i$th string at length $|x|$, $|q_r| = |x| + 1$, and $\langle q_1, q_2, \ldots, q_r\rangle \in C$. Thus, $\rho' = \{\langle x_1, x_2, \ldots, x_k\rangle \mid (x_1, x_2, \ldots, x_k) \in \rho\} \in \mathrm{P}^C$.

$\emptyset \leq_{\rho\text{-}T}^{p,C} B$ via a Turing machine $M$ that on input $x$ rejects if $x \notin D$. Since $(x) \in \rho$, $M$ has the $\rho$ query property with respect to $B$ on input $x$. Otherwise (that is, if $x \in D$), let $n = |x|$. By definition, $n = 2^{2^{2^{n_{k-1}}}}$, for some $k \in \mathbb{N}$. $M$ asks the following queries in sequence to its first oracle: $r_1, r_2, \ldots, r_{n+1}$, where $r_i$ is the lexicographically $i$th string of length $n$. Let $b_i$ be the answer that $M$ obtains from the first oracle to query $r_i$. $M$ now asks the query $b_1 b_2 \ldots b_{n+1}$ to its first oracle and rejects. Clearly, if $M$'s first oracle is $B$, then by construction at Stage $k$, $(x, r_1, r_2, \ldots, r_{n+1}, b_1 b_2 \ldots b_{n+1}) \in \rho_k$. Since $\rho_k \subseteq \rho$, it follows that $M$ has the $\rho$ query property with respect to $B$ on input $x$.

Assume that $\emptyset \leq_{r\text{-}\rho\text{-}T}^{p,C} B$ via $M_k$. We will derive a contradiction. Let $n = n_k$. First note that it follows from Lemma 3.6 that the behavior of $M^{B_k, C_k}(0^n)$ is exactly the same as that of $M^{B,C}(0^n)$. This is because in the stages after Stage $k$, the membership (in $B$ or in $C$) of strings up to length $n^k + k$ remains unaffected. Since Case 2 is impossible, Case 1 must hold in Stage $k$. Let $Y$ and $Y'$ be as defined in Case 1 of Stage $k$ of the construction. Let $M = M_k$. The sequence of queries that $M^{B_{k-1} \cup Y', C_k}(0^n)$ asks to its first oracle is not $q_Y$. Since each string queried by $M$ on input $0^n$ must have length at most $n^k + k$, the sequence of queries that $M^{B_{k-1} \cup Y', C}(0^n)$ asks to its first oracle is not $q_Y$. Thus, $M = M_k$ cannot show that $\emptyset \leq^{p,C}_{r\text{-}\rho\text{-}T} B$. Since $k$ was arbitrarily chosen, $\emptyset \not\leq^{p,C}_{r\text{-}\rho\text{-}T} B$. □

We mention in passing that Theorem 3.5 and many of the other results that we prove here via explicit diagonalization could be proved also via constructions that involve Kolmogorov complexity. Indeed, the November 2003 precursor of this paper [HT03a] does that for many of the results that it covers. However, as is often the case when doing Kolmogorov-based oracle constructions, the constructions involve stages, cases, and information counting that is sufficiently complex as to make the proofs long and in our opinion less easily understood than the explicit diagonalization versions included here.

## 4 Comparing Query-Monotonic Reductions to Turing and Truth-Table Reductions

In this section, we compare the power of query-monotonic reductions to different forms of Turing and truth-table reductions. By definition, query-monotonic reductions are no more powerful than Turing reductions. But in which cases are they as powerful as Turing reductions and in which cases are they not? How does the power of query-monotonic reductions relate to the that of classical reductions such as Turing reductions and truth-table reductions? These are the central questions that we answer in this section.

Let $\leq^p_\alpha$ and $\leq^p_\beta$ be defined reductions. If $(\forall A, B \subseteq \Sigma^*)[A \leq^p_\alpha B \implies A \leq^p_\beta B]$, then we say that $\leq^p_\beta$ *is stronger than* $\leq^p_\alpha$ and that $\leq^p_\alpha$ *is weaker than* $\leq^p_\beta$. If $\leq^p_\beta$ is stronger than $\leq^p_\alpha$ and $(\exists A, B \subseteq \Sigma^*)[A \leq^p_\beta B \land A \not\leq^p_\alpha B]$, then we say that $\leq^p_\beta$ *is strictly stronger than* $\leq^p_\alpha$ and that $\leq^p_\alpha$ *is strictly weaker than* $\leq^p_\beta$. (For example, $\leq^p_T$ is trivially stronger than itself, and is both stronger and strictly stronger than $\leq^p_{tt}$.)[4] If we say that $\leq^p_\beta$ is not stronger than $\leq^p_\alpha$, we mean just that—that "$\leq^p_\beta$ is stronger than $\leq^p_\alpha$" is untrue. That is, this means $(A, B \subseteq \Sigma^*)[A \leq^p_\alpha B \land A \not\leq^p_\beta B]$. Note of course that "$\leq^p_\beta$ is not stronger than $\leq^p_\alpha$" and "$\leq^p_\beta$ is strictly weaker than $\leq^p_\alpha$" are not synonymous. Also note that "$\leq^p_\beta$ is strictly stronger than $\leq^p_\alpha$" is equivalent to "$\leq^p_\beta$ is stronger than $\leq^p_\alpha$ but $\leq^p_\alpha$ is not stronger than $\leq^p_\beta$."

Let $X = \{li, ld, lni, lnd, s\text{-}li, s\text{-}ld, s\text{-}lni, s\text{-}lnd\}$. For each $\alpha, \beta \in X$, we ask whether $\leq^p_\alpha$ is stronger than $\leq^p_\beta$. There are 64 ($8 \times 8 = 64$) such questions. We resolve all these questions. In fact, we show that the only "stronger than" relationships that hold are the ones that immediately follow from the definitions. Since the definitions of query-monotonic reductions are based on restriction sets (see Definition 2.3), we can formally (yet succinctly) state the answers to each of these 64 questions as Theorem 4.1.

**Theorem 4.1** *For each $\alpha, \beta \in \{li, ld, lni, lnd, s\text{-}li, s\text{-}ld, s\text{-}lni, s\text{-}lnd\}$, $\leq^p_\beta$ is stronger than $\leq^p_\alpha$ if and only if $\rho_\alpha \subseteq \rho_\beta$.*

Note that, intuitively speaking, Theorem 4.1 says that the only stronger-than relationships that hold are the ones that are "obvious." Theorem 4.1 is a corollary of Theorems 4.9, 4.11, 4.13, and 4.14, all of which are stated and proven below.

---

[4]So, as is now standard in complexity, we use "stronger" to refer to reductions that link at least as many sets as the ones they are stronger than. To avoid confusion, we mention in passing that recursion theorists sometimes exchange "strong" and "weak," as they focus on the level of refinement of the equivalence classes induced by the reductions, and a few computer science papers—especially early ones—mirrored that notion.

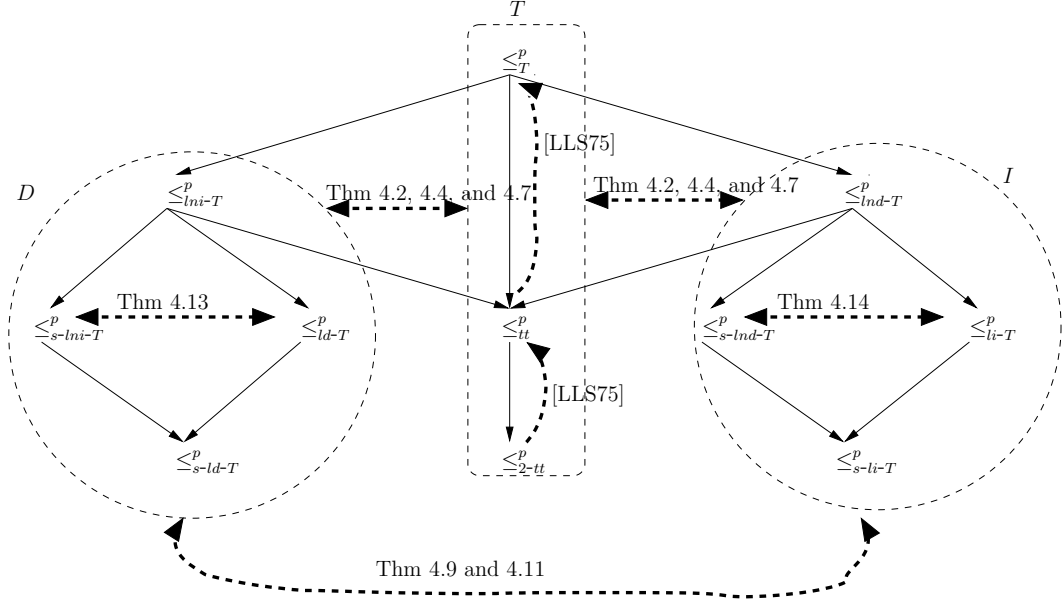

Figure 2: Stronger-than and not-stronger-than relationships between different query-monotonic reductions. Circles labeled $D$ and $I$ represent families of reductions (respectively, the "decreasing" family and the "increasing" family) defined in this paper. Box $T$ represents the "Turing/truth-table" family of reductions (namely, $\leq^p_T$, $\leq^p_{tt}$, and $\leq^p_{2\text{-}tt}$) that we compare our reductions against. A path of solid edges from $\leq^p_\beta$ to $\leq^p_\alpha$ indicates that $\leq^p_\beta$ is stronger than $\leq^p_\alpha$ (that is, $(\forall A, B \subseteq \Sigma^*)[A \leq^p_\alpha B \implies A \leq^p_\beta B]$). Note that all the solid edges are immediate from the definitions. A dashed edge from family of reductions $X$ to family of reductions $Y$ indicates that, for each $\leq^p_\beta \in X$ and $\leq^p_\alpha \in Y$, if there is no directed path of solid edges from $\leq^p_\beta$ to $\leq^p_\alpha$, then a solid edge from $\leq^p_\beta$ to $\leq^p_\alpha$ is impossible (in other words, $(\exists A, B \subseteq \Sigma^*)[A \leq^p_\alpha B \land A \not\leq^p_\beta B]$. A label on a dashed edge indicates the results from which the presence of the edge can be inferred.

We also want to compare query-monotonic reductions to classical notions of reductions (such as Turing reductions). Thus, for each $\alpha \in X$ and for each $\beta \in Y$, where $X$ is defined as above and $Y = \{T, tt, 2\text{-}tt\}$, we ask whether $\leq^p_\alpha$ is stronger than $\leq^p_\beta$ and we ask whether $\leq^p_\beta$ is stronger than $\leq^p_\alpha$. There are 48 $(2 \times 3 \times 8 = 48)$ such questions. We resolve each of these 48 questions in Theorems 4.2, 4.4, and 4.7.

Roughly speaking, it follows from Theorems 4.2, 4.4, 4.7, 4.9, 4.11, 4.13, and 4.14 that, for each $\alpha \in X$ and for each $\beta \in X \cup Y$, of the two stronger-than relationships that can hold between $\leq^p_\alpha$ than $\leq^p_\beta$, the only ones that hold are the ones that are *immediate* from the definitions. That is, if it is not immediate from the definitions that $\leq^p_\beta$ is stronger than $\leq^p_\alpha$, then $\leq^p_\beta$ is (provably) not stronger than $\leq^p_\alpha$. And, if two reductions *seem* incomparable, they are in fact (provably) incomparable.

The main results of this section (in particular, the answers to the questions mentioned above) are summarized in Figure 2. The solid directed edges in Figure 2 show the stronger-than relationships that are immediate from the definitions. The results of this section show that the only stronger-than relationships that hold are the ones shown by the solid edges. (Note that the stronger-than relation is a transitive relation. For clarity, Figure 2 only shows (using directed edges) a subset of the tuples in the stronger-than relation. The set of all tuples in the stronger-than relation is the transitive closure, under the stronger-than relation, of the set of tuples shown in Figure 2.) Thus, one way to look at the results of this section is: If there is no directed path (that uses only solid edges) between two arbitrary distinct nodes in Figure 2, no solid edge can be drawn between them. The dashed edges represent the not-stronger-than relationship. To avoid clutter, the nodes in Figure 2 have been clustered into 3 families: $D$ ("decreasing"), $I$ ("increasing"), and $T$ ("Turing/truth-table"). A dashed edge from $\leq_\beta^p$ to $\leq_\alpha^p$ indicates that $\leq_\beta^p$ is not stronger than $\leq_\alpha^p$. A dashed edge from a group $X \in \{D, I, T\}$ to a group $Y \in \{D, I, T\}$ indicates that, for each node $\leq_\beta^p \in X$ and $\leq_\alpha^p \in Y$, if there is no (solid) directed path from $\leq_\beta^p$ to $\leq_\alpha^p$, then $\leq_\beta^p$ is not stronger than $\leq_\alpha^p$. The label on each dashed edge denotes the set of results that imply the not-stronger-than relationships that the edge denotes.

Given reductions $\leq_\alpha^p$ and $\leq_\beta^p$ such that there is no (solid) directed path from $\leq_\beta^p$ to $\leq_\alpha^p$ in Figure 2, one can also use Figure 2 to derive the result that $\leq_\beta^p$ is not stronger than $\leq_\alpha^p$. To show that $\leq_\beta^p$ is not stronger than $\leq_\alpha^p$, it is sufficient to show that adding a solid (directed) edge from $\leq_\beta^p$ to $\leq_\alpha^p$ in Figure 2 results in the following situation: Let $G$ be the graph shown in Figure 2 and let $G'$ be $G$ appended with the edge $(\leq_\beta^p, \beta_\alpha^p)$. There are distinct nodes $\leq_\gamma^p$ and $\leq_\delta^p$ in $G$ such that there is a dashed edge from $\leq_\gamma^p$ to $\leq_\delta^p$ in $G$, there is no directed path from $\leq_\gamma^p$ to $\leq_\delta^p$ in $G$, yet in $G'$ there is a (solid) directed path from $\leq_\gamma^p$ to $\leq_\delta^p$. This a contradiction because solid edges represent stronger-than relationships while dashed edges represent not-stronger-than relationships. (Note that the stronger-than relation is transitive.)

Thus, Figure 2 can be used as a ready reference for comparing query-monotonic reductions. For example, to look up the known relationships between, say, strong query-nonincreasing reductions and query-increasing reductions, first locate $\leq_{s\text{-}lni\text{-}T}^p$ and $\leq_{li\text{-}T}^p$ in Figure 2 and find if there is any (solid) directed path between these nodes. In this case there is none. This means that $\leq_{s\text{-}lni\text{-}T}^p$ is incomparable to $\leq_{li\text{-}T}^p$. Proofs for not-stronger-than relationships between $\leq_{s\text{-}lni\text{-}T}^p$ and $\leq_{li\text{-}T}^p$ can be obtained as follows. Find the dashed path between circles $D$ (which contains $\leq_{s\text{-}lni\text{-}T}^p$) and $I$ (which contains $\leq_{li\text{-}T}^p$). The path is labeled "Thm 4.9 and 4.11." This means that using one or more of Theorems 4.9 and 4.11, we can prove that $\leq_{s\text{-}lni\text{-}T}^p$ and $\leq_{li\text{-}T}^p$ are in fact incomparable. It turns out that we need both Theorem 4.9 and Theorem 4.11. To show that $\leq_{li\text{-}T}^p$ is not stronger than $\leq_{s\text{-}lni\text{-}T}^p$ (in other words, to show that there exist $A, B \subseteq \Sigma^*$ such that $A \leq_{s\text{-}lni\text{-}T}^p B$ yet $A \not\leq_{li\text{-}T}^p B$), we use Theorem 4.9 (which says that $\leq_{lnd\text{-}T}^p$ is not stronger than $\leq_{s\text{-}ld\text{-}T}^p$) along with the observations: $\leq_{lnd\text{-}T}^p$ is stronger than $\leq_{li\text{-}T}^p$ and $\leq_{s\text{-}lni\text{-}T}^p$ is stronger than $\leq_{s\text{-}ld\text{-}T}^p$. (These two results can be inferred from the presence of directed paths from $\leq_{lnd\text{-}T}^p$ to $\leq_{li\text{-}T}^p$ and from $\leq_{s\text{-}lni\text{-}T}^p$ to $\leq_{ld\text{-}T}^p$, respectively.) Note that if we assume (for the purpose of deriving a contradiction) that $\leq_{li\text{-}T}^p$ is stronger than $\leq_{s\text{-}lni\text{-}T}^p$, then by the transitivity of the stronger-than relation we have that $\leq_{lnd\text{-}T}^p$ is stronger than $\leq_{s\text{-}ld\text{-}T}^p$, which is a contradiction to Theorem 4.9. To show that $\leq_{s\text{-}lni\text{-}T}^p$ is not stronger than $\leq_{li\text{-}T}^p$, we combine the result that query-increasing reductions are not stronger than strong query-increasing reductions (Theorem 4.11) with the observations: $\leq_{lni\text{-}T}^p$ is stronger than $\leq_{s\text{-}lni\text{-}T}^p$ and $\leq_{li\text{-}T}^p$ is stronger than $\leq_{s\text{-}li\text{-}T}^p$.

We will now state and prove results from which all the not-stronger-than relationships can be derived. We first state (as Theorems 4.2, 4.4, and 4.7) the not-stronger-than relationships between the query-monotonic reductions and the classical reductions. This set of results corresponds to the bidirectional dashed arrows between $D$ and $T$ and between $I$ and $T$ in Figure 2. After this set of results, we will state (as Theorems 4.9, 4.11, 4.13, and 4.14) the not-stronger-than relationships between query-monotonic reductions. This set of results corresponds to the bidirectional dashed edges between $D$ and $I$, between $\leq_{s\text{-}lni\text{-}T}^p$ and $\leq_{ld\text{-}T}^p$, and between $\leq_{s\text{-}lnd\text{-}T}^p$ and $\leq_{li\text{-}T}^p$.

In Theorem 4.2, we show that two strongest query-monotonic reductions (namely, query-nonincreasing reductions and query-nondecreasing reductions) are not stronger than the Turing reductions. Given that Turing reductions are stronger than both query-nonincreasing and query-nondecreasing reductions, Theorem 4.2 implies that Turing reductions are *strictly* stronger than both query-nonincreasing and query-nondecreasing reductions.

**Theorem 4.2** $(\exists A, B \subseteq \Sigma^*)[B \leq_T^p A \wedge B \not\leq_{lni\text{-}T}^p A \wedge B \not\leq_{lnd\text{-}T}^p A]$.

**Proof** Intuitively speaking, the proof is based on the observation that a Turing machine implementing a query-nonincreasing reduction cannot ask a longer query after it has asked a shorter one. Similarly, a machine implementing a query-nondecreasing reduction cannot ask a shorter query after it has asked a longer one. The diagonalization works by ensuring that for each machine $M$ there is a diagonalizing string $x$ such that the membership (in the test language[5] $B = L_A$) of $x$ depends on two strings in the oracle $A$; let us call them $\ell$ (for "long") and $s$ (for "short"). $\ell$ has length longer than $|x|$ and $s$ has length shorter than $|x|$. We will ensure that $\ell$ and $s$ are picked so that if $M$ asks queries in a length-nonincreasing manner, it cannot query $\ell$ to its oracle and if $M^A(x)$ asks query in a length-nondecreasing manner, it cannot query $s$ to its oracle. Since the membership of $x$ will depend on both $\ell$ and $s$, we can then force $M$ to have the behavior that we need for diagonalization.

For each $D \subseteq \Sigma^*$, we define $L_D$ as:

$$L_D = \{0^{4^k} \mid k \geq 1 \wedge (\mu_1(k) \in D) \oplus (\mu_2(k) \in D))\},$$

where, for each $k \in \mathbb{N}^+$, $\mu_1(k)$ is the string $0^{(1/4)4^k}\chi_D(0^{4^k-1})\chi_D(0^{4^k-2})\ldots\chi_D(0^{(3/4)4^k})$ and $\mu_2(k)$ is the string $0^{(5/4)4^k}\chi_D(0^{4^k+1})\chi_D(0^{4^k+2})\ldots\chi_D(0^{(5/4)4^k})$. Note that for each $D \subseteq \Sigma^*$, $L_D \leq_T^p D$. We will construct the set $A$ in stages below such that $L_A \not\leq_{lni\text{-}T}^p A$ and $L_A \not\leq_{lnd\text{-}T}^p A$. Thus, $L_A$ will be the set $B$ promised in the statement of this theorem. In odd numbered stages we will ensure that $L_A \not\leq_{lni\text{-}T}^p A$ and in even numbered stages we will ensure that $L_A \not\leq_{lnd\text{-}T}^p A$. For each $i \in \mathbb{N}^+$, let $A_{i-1}$ denote the set of strings added to $A$ before Stage $i$. $A_0 = \emptyset$.

**Stage $i = 2j - 1$, $j \in \mathbb{N}^+$.** Let $M = M_j$. We will ensure that $M$ does not implement $L_A \leq_{lni\text{-}T}^p A$. Let $n = n_i$ be the smallest integer such that no string of length greater than or equal to $n$ has been made relevant[6] in any previous stage, $n^i + i < 2^{(1/4)n}$, and for some $k \geq 1$, $n = 4^k$. If $M^{A_{i-1}}(0^{4^k})$ does not ask queries in a length-nonincreasing manner, let $A_i = A_{i-1}$. Go to Stage $i+1$. If $M^{A_{i-1}}(0^{4^k})$ asks queries in a length-nonincreasing manner, let $\alpha$ be a string of length $(1/4)4^k$ such that $M^{A_{i-1}}(0^{4^k})$ does not query $0^{(5/4)4^k}\alpha$ (for specificity, the lexicographically least such string). (Such a string exists because $n^i + i < 2^{(1/4)4^k}$ by our choice of $n$.) We will fix the membership (in $A$) of the following strings in accordance with $\alpha$: $0^{4^k+1}$, $0^{4^k+2}$, ..., $0^{(5/4)4^k}$. In particular, for each $1 \leq i \leq (1/4)4^k$, if the $i$th bit of $\alpha$ is 0, we put the string $0^{4^k+i}$ out of $A$ and if the $i$th bit of $\alpha$ is 1, we put the string $0^{4^k+i}$ in $A$. Let $A'$ be resulting set. Formally, let $A' = A_{i-1} \cup X$, where $X = \{0^{4^k+i} \mid 1 \leq i \leq (1/4)4^k \text{ and } i\text{th bit of } \alpha \text{ is } 1\}$. Since, by assumption, $M^{A_{i-1}}(0^{4^k})$ does not query $0^{(5/4)4^k}\alpha$, if $M^{A'}(0^{4^k})$ queries $0^{(5/4)4^k}\alpha$, then $M^{A'}(0^{4^k})$ queries $0^{(5/4)4^k}\alpha$ after having queried

---

[5] The term "test language" doesn't have any formal, technical meaning, but rather is typically used in construction settings to describe a language that is defined in terms of another one—in this case, $B$ is defined via $A$ in a manner that will soon be specified—and that typically is used to instantiate or falsify some reducibility or membership claim (see, e.g., [Tha04]).

[6] If a stage of this construction speaks of a machine, $M_k$, running on inputs $x_1, x_2 \ldots$, then we say that all strings of length up to and including

$$\max_{x_i}(|x_i|^k + k)$$

are made *relevant* by this stage (regardless of whether or not they are queried), and we also say that each string that, in that stage, we explicitly place into the oracle or explicitly set as not belonging to the oracle is made relevant by this stage. (For later proofs, we will not as explicitly define "relevant," but it will in each case be analogous, and clear from content.)

some string from the set $X$ (since $A_{i-1}$ and $A$ differ only regarding $X$). Each string in $X$ has length strictly less than $|0^{(5/4)4^k}\alpha|$. Thus, in this case $M$ does not have the query-nonincreasing property with respect to $A'$ on input $0^{4^k}$. Let $A_i = A'$. Go to Stage $i+1$. Otherwise (that is, if $M^{A'}(0^{4^k})$ does not query $0^{(5/4)4^k}\alpha$), define $A_i = A' \cup \{0^{(5/4)4^k}\alpha \mid M^{A'}(0^{4^k})$ rejects$\}$.[7] Clearly, $M^{A_i}(0^{4^k})$ accepts if and only if $0^{4^k} \notin L_{A_i}$. We will maintain this behavior throughout the construction.[8] Go to Stage $i+1$.

**End of Stage $i$.**

**Stage $i = 2j$, $j \in \mathbb{N}^+$.** Let $M = M_j$. We will ensure that $M$ does not implement $L_A \leq^p_{lnd\text{-}T} A$. Let $n = n_i$ be the smallest integer such that no string of length greater than or equal to $n$ has been made relevant in any previous stage, $n^i + i < 2^{(1/4)n}$, and for some $k \geq 1$, $n = 4^k$. If $M^{A_{i-1}}(0^{4^k})$ does not ask queries in a length-nondecreasing manner, let $A_i = A_{i-1}$. Go to Stage $i+1$. If $M^{A_{i-1}}(0^{4^k})$ asks queries in a length-nondecreasing manner, let $\alpha$ be a string of length $(1/4)4^k$ such that $M^{A_{i-1}}(0^{4^k})$ does not query $0^{(1/4)4^k}\alpha$ (for specificity, the lexicographically least such string). (Such a string exists because $n^i + i < 2^{(1/4)4^k}$ by our choice of $n$.) We will fix the membership (in $A$) of the following strings in accordance with $\alpha$: $0^{4^k-1}$, $0^{4^k-2}$, ..., $0^{(3/4)4^k}$. In particular, for each $1 \leq i \leq (1/4)4^k$, if the $i$th bit of $\alpha$ is 0, we put the string $0^{4^k-i}$ out of $A$ and if the $i$th bit of $\alpha$ is 1, we put the string $0^{4^k-i}$ in $A$. Let $A'$ be resulting set. Formally, let $A' = A_{i-1} \cup X$, where $X = \{0^{4^k-i} \mid 1 \leq i \leq (1/4)4^k$ and $i$th bit of $\alpha$ is 1$\}$. Since, by assumption, $M^{A_{i-1}}(0^{4^k})$ does not query $0^{(1/4)4^k}\alpha$, if $M^{A'}(0^{4^k})$ queries $0^{(1/4)4^k}\alpha$, then $M^{A'}(0^{4^k})$ queries $0^{(1/4)4^k}\alpha$ after having queried some string from the set $X$. Each string in $X$ has length strictly greater than $|0^{(1/4)4^k}\alpha|$. Thus, in this case $M$ does not have the query-nondecreasing property with respect to $A'$ on input $0^{4^k}$. Let $A_i = A'$. Go to Stage $i+1$. Otherwise (that is, if $M^{A'}(0^{4^k})$ does not query $0^{(1/4)4^k}\alpha$), define $A_i = A' \cup \{0^{(1/4)4^k}\alpha \mid M^{A'}(0^{4^k})$ rejects$\}$. Clearly, $M^{A_i}(0^{4^k})$ accepts if and only if $0^{4^k} \notin L_{A_i}$. We will maintain this behavior throughout the construction. Go to Stage $i+1$.

**End of Stage $i$.**

Let $A = \bigcup_{i \geq 1} A_i$. It is clear from our construction in the odd numbered stages that, for each $j \in \mathbb{N}^+$, $M_j$ does not implement $L_A \leq^p_{lni\text{-}T} A$. Thus, $L_A \not\leq^p_{lni\text{-}T} A$. Furthermore, it is clear from our construction in the even numbered stages that, for each $j \in \mathbb{N}^+$, $M_j$ does not implement $L_A \leq^p_{lnd\text{-}T} A$. Thus, $L_A \not\leq^p_{lnd\text{-}T} A$. ❑

As an immediate corollary to Theorem 4.4, we get the following.

**Corollary 4.3** $(\exists A, B \subseteq \Sigma^*)[B \leq^p_T A \wedge B \not\leq^p_{li\text{-}T} A \wedge B \not\leq^p_{ld\text{-}T} A]$.

Roughly speaking, in Theorems 4.4, 4.6, and 4.7 we show that nonadaptive Turing reductions (i.e., truth-table reductions) and query-increasing Turing reductions are incomparable: Theorem 4.4 shows that neither query-increasing nor query-decreasing Turing reductions are stronger than two-truth-table reductions (and thus, neither is stronger than truth-table reductions). Nonetheless, we show in Theorem 4.7 that truth-table reductions are not stronger than either query-increasing or query-decreasing Turing reductions. In fact, we show in Theorem 4.7 that truth-table reductions are not stronger than even strong query-increasing Turing reductions and truth-table reductions are not stronger than even query-decreasing Turing reductions. Note that Theorem 4.4 as an immediate corollary implies that neither query-increasing nor query-decreasing reductions are stronger than Turing reductions.

---

[7]We will often use this notation. It means just what it says interpreted in the natural mathematical way: $A_i = A' \cup \{0^{(5/4)4^k}\alpha\}$ if $M^{A'}(0^{4^k})$ rejects, and otherwise $A_i = A'$.

[8]The sentence "We will maintain this behavior throughout the construction," which will appear often in this paper, means that the behavior just mentioned—in this particular case "$M^{A_i}(0^{4^k})$ accepts if and only if $0^{4^k} \notin L_{A_i}$"—will hold not just for $A_i$ but even for $A$, i.e., no action taken later in the construction will break this.

**Theorem 4.4** $(\exists A, B \subseteq \Sigma^*)[B \leq_{2\text{-}tt}^p A \land B \not\leq_{li\text{-}T}^p A \land B \not\leq_{ld\text{-}T}^p A]$.

**Proof** Roughly speaking, the proof is based on exploiting the "maximum of one query per length" limitation of machines implementing query-increasing and query-decreasing reductions. In contrast to this limitation, even though a machine implementing a 2-truth-table reduction is allowed at most two queries to its oracles on any given input, both these queries may be of the same length. We exploit this difference in our diagonalization to construct sets $A$ and $B = L_A$ such that $L_A \leq_{2\text{-}tt}^p A$, yet $L_A \not\leq_{li\text{-}T}^p A$ and $L_A \not\leq_{ld\text{-}T}^p A$. For each $D \subseteq \Sigma^*$, define $L_D$ as follows:

$$L_D = \{0^i \mid (0^i \in D) \oplus (1^i \in D)\},$$

where $\oplus$ is the exclusive OR operator. Clearly, for each $D$, $L_D \in \mathrm{R}_{2\text{-}tt}^p(D)$. We will now construct a set $A$ such that $L_A \notin \mathrm{R}_{li\text{-}T}^p(A) \cup \mathrm{R}_{ld\text{-}T}^p(A)$. We will construct $A$ in stages by adding zero or more elements to $A$ at each stage. At each stage, we will ensure that the elements added to $A$ at that stage have not been made relevant (see footnote 6) in earlier stages. For each $k$, $n_k \in \mathbb{N}$ (defined later) is the only length at which strings are added at Stage $k$. Let $A_{k-1}$ be the set consisting of all the elements added to $A$ before Stage $k$. In Stage $k$, we will ensure, via diagonalization if required, that at least one of the following two conditions hold:

1. $M_i$ neither has the $\rho_{li}$ query property nor has the $\rho_{ld}$ query property with respect to $A$.

2. $L(M_i^A) \neq L_A$.

Note that if, for each $k$, at least one of the above conditions hold, then $L_A \not\leq_{li\text{-}T}^p A$ and $L_A \not\leq_{ld\text{-}T}^p A$.

Let $A_0 = \emptyset$ and $n_0 = 1$.

**Stage $k \in \mathbb{N}^+$.** Choose $n = n_k$ to be the smallest natural number that is is larger than the length of all strings made relevant (see footnote 6) in earlier stages. Let $M = M_k$. Either $M^{A_{k-1}}(0^n)$ accepts or $M^{A_{k-1}}(0^n)$ rejects. If $M^{A_{k-1}}(0^n)$ accepts, then let $A_k = A_{k-1}$. $M^{A_k}(0^n)$ accepts, yet $0^n \notin L_{A_k}$, and so condition 2 is satisfied. We will maintain this behavior throughout the construction. If $M^{A_{k-1}}(0^n)$ rejects, then there are 3 possible cases:

1. $M^{A_{k-1}}(0^n)$ queries neither $0^n$ nor $1^n$.

2. $M^{A_{k-1}}(0^n)$ queries exactly one of $0^n$ and $1^n$.

3. $M^{A_{k-1}}(0^n)$ queries both $0^n$ and $1^n$.

If $M^{A_{k-1}}(0^n)$ queries neither $0^n$ nor $1^n$, then let $A_k = A_{k-1} \cup \{0^n\}$. Clearly, $0^n \in L_{A_k} - L(M^{A_k})$. We will maintain this behavior throughout the construction. Thus, condition 2 is satisfied. If $M^{A_{k-1}}(0^n)$ queries exactly one of $0^n$ and $1^n$, we let $A_k = A_{k-1} \cup \{a\}$, where $a$ is the string in $\{0^n, 1^n\}$ that is *not* queried by $M^{A_{k-1}}(0^n)$. Clearly, $0^n \in L_{A_k} - L(M^{A_k})$. We will preserve this behavior throughout the construction. Thus, condition 2 is satisfied. If $M^{A_{k-1}}(0^n)$ queries both $0^n$ and $1^n$, then let $A_k = A_{k-1}$. In this case, since $M$ queries two strings of the same length, $M$ neither has the $\rho_{li}$ query property nor has the $\rho_{ld}$ query property with respect to $A_k$. We will maintain this behavior throughout the construction. Thus, condition 1 is satisfied. Go to Stage $k + 1$.
**End of Stage $k$.**

Let $A = \bigcup_{k \geq 1} A_k$. As noted earlier, if for each $k$, at least one of the two conditions holds, then $L_A \not\leq_{li\text{-}T}^p A$ and $L_A \not\leq_{ld\text{-}T}^p A$. It is evident from the construction that, for each $k \geq 1$, at least one of the two conditions holds in Stage $k$ of the construction. Also, by our choice of $n_k$ in subsequent stages, the behavior that causes this condition to hold at Stage $k$ is preserved in subsequent stages, and thus this condition itself continues to hold in subsequent stages. ❑

Theorem 4.4 shows that polynomial-time query-increasing Turing reductions and polynomial-time 2-truth-table reductions differ. In fact, the proof of Theorem 4.4 actually achieves a slightly stronger result. Note that the $L_A$ we construct in the proof not only polynomial-time 2-truth-table reduces to $A$, but also 2-truth-table reduces to $A$ via a machine that uses the *same* truth-table (namely, the "2-ary parity" truth-table) for each input.

**Definition 4.5** *[Wec85] A $k$-fixed-truth-table reduction, denoted $\leq^p_{k\text{-}ftt}$, is a truth-table reduction in which the machine implementing the reduction uses the same truth-table for each input.*

There are a total of sixteen two-argument fixed-truth-table reductions (corresponding to the sixteen 2-ary boolean connectives). A *completely degenerate* boolean connective is one whose output does not depend on any of its inputs, and an *almost-completely degenerate* boolean connective is one whose output depends on exactly one of its inputs ([HJ95], which shows that these notions are highly relevant to closure properties of the P-selective sets). Of the $2^{2^k}$ $k$-ary boolean connectives, there are exactly two completely degenerate ones and exactly $2k$ almost-completely degenerate ones. Since each boolean connective corresponds to a $k$-fixed-truth-table reduction, there are two completely degenerate $k$-fixed-truth-table reductions and $2k$ almost-completely degenerate $k$-fixed-truth-table reductions. For each $k$-ary boolean connective $\sigma$, let $\leq^p_{\sigma\text{-}ftt}$ denote the $k$-fixed-truth-table reduction in which the machine uses $\sigma$ as its truth-table.

As observed earlier, the test language $L_A$ in the proof of Theorem 4.4 is not only in $\mathrm{R}^p_{2\text{-}tt}(A)$, as claimed in the statement of theorem, but also in $\mathrm{R}^p_{\sigma_\oplus\text{-}ftt}(A)$, where $\sigma_\oplus$ is the "2-ary parity" truth-table. It is not hard to see that for each 2-ary boolean connective $\sigma$ that is neither completely degenerate nor almost-completely degenerate, the construction of Theorem 4.4 can be modified (the "3 possible cases"/subsequent arguments framework changes in some of these cases, but in each case it is not hard to see that diagonalization can be successfully achieved) so that the test language $L_A \leq^p_{\sigma\text{-}ftt} A$ yet $L_A \not\leq^p_{li\text{-}T} A$ and $L_A \not\leq^p_{ld\text{-}T} A$. On the other hand, if $\sigma$ is completely degenerate or almost-completely degenerate then, for each $B \in \mathrm{R}^p_{\sigma\text{-}ftt}(A)$, the membership of any string in $B$ depends on the membership (in $A$) of at most one string. Thus, $B \leq^p_{li\text{-}T} A$ and $B \leq^p_{ld\text{-}T} A$. We formalize these relationships between fixed-truth-table reductions and query-monotonic reductions in Theorem 4.6 below.

**Theorem 4.6**  1. *For each 2-ary boolean connective $\sigma$ that is neither completely degenerate nor almost-completely degenerate, there exists an $A, B \subseteq \Sigma^*$ such that $B \leq^p_{\sigma\text{-}ftt} A$ yet $B \not\leq^p_{li\text{-}T} A$ and $B \not\leq^p_{ld\text{-}T} A$.*

2. *For each 2-ary boolean connective $\sigma$ that is completely degenerate or almost-completely degenerate, and for each $A, B \subseteq \Sigma^*$, $B \leq^p_{\sigma\text{-}ftt} A \implies (B \leq^p_{li\text{-}T} A \wedge B \leq^p_{ld\text{-}T} A)$.*

Theorems 4.4 and 4.6 show that neither query-increasing nor query-decreasing reductions are stronger than truth-table reductions. In fact, neither query-increasing nor query-decreasing reductions are stronger than even 2-query fixed-truth-table reductions. In light of these results, it is interesting to ask the converse question: Are truth-table reductions stronger than query-increasing reductions? Are truth-table reductions stronger than query-decreasing reductions? In Theorem 4.7 we show that the answer to both these questions is "no."

**Theorem 4.7** $(\exists A, B \subseteq \Sigma^*)[B \leq^p_{s\text{-}li\text{-}T} A \wedge B \leq^p_{s\text{-}ld\text{-}T} A \wedge B \not\leq^p_{tt} A]$.

**Proof** The proof is based on the construction of an oracle $A$ and a test language $B$ such that membership of strings in $B$ can be decided in two different ways—by querying $A$ in a strong length-increasing fashion or by querying $A$ in a strong length-decreasing fashion. Additionally, we ensure via diagonalization that $B$ does not polynomial-time truth-table reduce to $A$.

For each $D$, define $L_D$ as follows:

$$L_D = \{0^{8^k} \mid k \geq 1 \wedge 0^{(5/8)8^k}\chi_D(0^{8^k-1})\chi_D(0^{8^k-2})\ldots\chi_D(0^{(7/8)8^k}) \in D\}.$$

We will construct sets $A$ and $L_A$ (which will be our $B$) such that $L_A \leq^p_{\text{s-li-}T} A$ and $L_A \leq^p_{\text{s-ld-}T} A$, yet $L_A \not\leq^p_{tt} A$.

Note that, for each $D \subseteq \Sigma^*$, $L_D \leq^p_{\text{s-ld-}T} D$ via a Turing machine $M$ that rejects any input that is not of the form $0^{8^k}$ and that, for any input of the form $0^{8^k}$, asks its oracle the following strings in order: $0^{8^k-1}$, $0^{8^k-2}$, ..., $0^{(7/8)8^k}$. From the answers $b_1, b_2, \ldots, b_{(1/8)8^k}$ to these queries, $M$ computes the string $q = 0^{(5/8)8^k}b_1b_2\ldots b_{(1/8)8^k}$, asks $q$ to its oracle, and accepts exactly if the oracle says "yes." Since $|q| = (5/8)8^k + (1/8)8^k = (3/4)8^k < (7/8)8^k$, it follows that $M$ indeed has the strong query-decreasing property with respect to each oracle $D$.

To ensure that the $L_A$ we construct strong query-increasing reduces to $A$, we will ensure that set $A$ that we construct obeys the following promises:

1. For each $k \geq 1$ and for each $j$ such that $1 \leq j \leq (1/8)8^k$, $\chi_A(0^{8^k-j}) = \chi_A(0^{8^k+j})$.

2. For each $k \geq 1$, $0^{(5/8)8^k}\chi_A(0^{8^k-1})\chi_A(0^{8^k-2})\ldots\chi_A(0^{(7/8)8^k}) \in A$ if and only if $0^{(9/8)8^k}\chi_A(0^{8^k+1})\chi_A(0^{8^k+2})\ldots\chi_A(0^{(9/8)8^k}) \in A$.

If these promises are satisfied, then $L_A \leq^p_{\text{s-li-}T} A$ via a Turing machine $M$ that rejects any input that is not of the form $0^{8^k}$ and that, on any input of the form $0^{8^k}$, asks its oracle the following $(1/8)8^k$ strings in order: $0^{8^k+1}$, $0^{8^k+2}$, ..., $0^{(9/8)8^k}$. From part 1 of the promises above, it follows that if the oracle providing the answers is $A$, then the sequence of answers to the queries above is exactly the sequence of answers to the following sequence of queries: $0^{8^k-1}$, $0^{8^k-2}$, ..., $0^{(7/8)8^k}$. Let this sequence be $b_1, b_2, \ldots, b_{(1/8)8^k}$. $M$ computes the string $q = 0^{(9/8)8^k}b_1b_2\ldots b_{(1/8)8^k}$, asks $q$ to its oracle, and accepts exactly if the oracle says "yes." From part 2 of the promises above, it follows that $q \in A$ if and only if $0^{(5/8)8^k}\chi_A(0^{8^k-1})\chi_A(0^{8^k-2})\ldots\chi_A(0^{(7/8)8^k}) \in A$. Thus, it follows from the definition of $L_A$ that, for each $k \geq 1$, $M(0^{8^k})$ accepts if and only if $0^{8^k} \in L_A$. It is immediate from its description that $M$ has the strong query-increasing property with respect to $A$ (and, in fact, with respect to each oracle).

It is not hard to see that each truth-table reduction can be implemented by some Turing machine $T$ such that, for each input $x$ and for each oracle $X$, in spirit "$T^X(x)$ computes all its oracle queries before asking any one of them." Of course, this is not a well-formalized notion (short of changing the model of what a Turing reduction itself is). But we can formalize the closely related notion that we want as follows. Instead of having "$T^X(x)$ computes all its oracle queries before asking any one of them" as our requirement, we will have as our requirement that for each $x$ and each $Y$ and each $Z$, the ordered sequence of queries asked by $T^Y(x)$ is exactly the same as the ordered sequence of queries asked by $T^Z(x)$ (this in effect ensures that the vector of questions asked by $T(x)$ does not depend on what our oracle is). Let us henceforward call Turing machines that have this property "truth-table machines." Let $T_1, T_2, \ldots$ be an (easily computed) enumeration of truth-table machines such that, for each $X \subseteq \Sigma^*$ and each $x \in \Sigma^*$, $T_i^X(x)$ runs in time $n^i + i$, and such that each truth-table reduction is implemented by at least one $T_i$ (note that we do not require that every machine with "truth-table machine" behavior appears in our enumeration, as that would run aground on undecidability issues; rather, we just require that our enumeration holds at least one machine implementing each truth-table reduction). It is not hard to see that such an enumeration exists, and we will from now on refer to it when we speak of $T_1, T_2, \ldots$. (Note that this paragraph presented a somewhat different approach to capturing truth-table reductions than that part 1 of Definition 2.1.)

We will now describe our construction of $A$. We will construct $A$ in stages. In Stage $i \geq 1$, we will diagonalize against the machine $T_i$. That is, in Stage $i$ we will show that $T_i$ does not implement $L_A \leq_{tt}^p A$.

Let $A_{i-1}$ denote the set of strings added to $A$ before Stage $i$. Let $A_0 = \emptyset$.

**Stage $i$, $i \geq 1$**: Let $n = n_i$ be the smallest integer such that, for some $k \geq 1$, $n = 8^k$, $n^i + i < 2^{n/8}$, and no string of length $(3/4)8^k$ or longer has been made relevant (see footnote 6) in any earlier stage.

We will diagonalize against $T_i$. Let $M = T_i$. Let $Q$ be the set of oracle queries that $M$ asks on input $0^{8^k}$. Since $M$ is a truth-table machine, the set of oracle queries does not depend on the oracle attached to $M$. Since $M$ runs in time at most $n^i + i$, $\|Q\| \leq n^i + i$. Let $\alpha$ be the lexicographically smallest string of length $n/8$ such that neither $0^{(5/8)8^k}\alpha$ nor $0^{(9/8)8^k}\alpha$ is in $Q$. Since $n^i + i < 2^{(1/8)8^k}$, $\alpha$ is well defined. We will fix the membership (in $A$) of the following strings in accordance with $\alpha$: $0^{8^k-1}, 0^{8^k-2}, \ldots, 0^{(7/8)8^k}$. In particular, for each $1 \leq i \leq (1/8)8^k$, if the $i$th bit of $\alpha$ is 0, we fix the string $0^{8^k-i}$ to be out of $A$ and if the $i$th bit of $\alpha$ is 1, we fix the string $0^{8^k-i}$ to be in $A$. Since we want $A$ to obey the promises, we make sure that $0^{8^k-i}$ is in $A$ exactly if $0^{8^k+i}$ is in $A$. Let $A'$ be resulting set. Formally, let $A' = A_{i-1} \cup \{0^{8^k-i} \mid 1 \leq i \leq (1/8)8^k$ and the $i$th bit of $\alpha$ is $1\} \cup \{0^{8^k+i} \mid 1 \leq i \leq (1/8)8^k$ and the $i$th bit of $\alpha$ is $1\}$. Define $A_i = A' \cup \{0^{(5/8)8^k}\alpha \mid M^{A'}(0^{8^k})$ rejects$\} \cup \{0^{(9/8)8^k}\alpha \mid M^{A'}(0^{8^k})$ rejects$\}$.

First note that if $A_{i-1}$ satisfies the abovementioned promises, then so does $A_i$. By the definition of $\alpha$, $M^{A_i}(0^{8^k})$ queries neither $0^{(5/8)8^k}\alpha$ nor $0^{(9/8)8^k}\alpha$. Thus, $M^{A_i}(0^{8^k})$ accepts if and only if $M^{A'}(0^{8^k})$ accepts, and so $M^{A_i}(0^{8^k})$ accepts if and only if $0^{8^k} \notin L_{A_i}$. We will maintain this behavior throughout the construction. Go to Stage $i+1$.

**End of Stage $i$**.

Let $A = \bigcup_{i \geq 1} A_i$. Since $A_0 = \emptyset$ trivially satisfies the promises, it follows from the construction that $A$ satisfies the promises. Thus, it is clear from the construction that $L_A \leq_{\text{s-li-}T}^p A$ and $L_A \leq_{\text{s-ld-}T}^p A$, yet $L_A \not\leq_{tt}^p A$. ❑

As an immediate corollary we have the following:

**Corollary 4.8**    *1. $(\exists A, B \subseteq \Sigma^*)[B \leq_{\text{li-}T}^p A \wedge B \leq_{\text{ld-}T}^p A \wedge B \not\leq_{tt}^p A]$.*

*2. $(\exists A, B \subseteq \Sigma^*)[B \leq_{\text{lni-}T}^p A \wedge B \leq_{\text{lnd-}T}^p A \wedge B \not\leq_{tt}^p A]$.*

In Theorems 4.2, 4.4, and 4.7, we saw not-stronger-than relationships between query-monotonic reductions and classical reductions (such as Turing and truth-table reductions). We now state and prove (as Theorems 4.9, 4.11, 4.13, and 4.14) not-stronger-than relationships between different query-monotonic reductions.

Note that truth-table reductions are at least as restrictive as query-nondecreasing (query-nonincreasing) reductions, because if all the queries are generated before any query is asked, then they can be asked in query-nondecreasing or query-nonincreasing order, regardless of what the queries are. Thus, for sets $A$ and $B$, if $A \leq_{tt}^p B$, then $A \leq_{\text{lni-}T}^p B$ and $A \leq_{\text{lnd-}T}^p B$. On the other hand, if two of the queries in the list of queries are of the same length, then they cannot be asked in query-increasing or in query-decreasing order. Thus, it is not clear whether truth-table reductions are as restrictive as query-increasing (query-decreasing) reductions. In fact, as Theorem 4.4 and Theorem 4.7 show, truth-table and query-increasing (query-decreasing) reductions are incomparable. In light of these results, it is interesting to ask how query-increasing and query-decreasing Turing reductions compare. Is one stronger than the other? We show in Corollaries 4.10 and 4.12 that query-increasing and query-decreasing Turing reductions are strength-wise incomparable. In fact, we prove the following stronger results in Theorems 4.9 and 4.11 from which Corollaries 4.10 and 4.12 follow easily: There exist sets $A$ and $B$ such that $B$ strong query-decreasing Turing reduces to $A$ but $B$ does not query-nondecreasing Turing reduce to $A$, and there also exist sets $A$ and $B$ such that $B$ strong query-increasing Turing reduces to $A$, yet $B$ does not query-nonincreasing Turing reduce to $A$.

**Theorem 4.9** $(\exists A, B \subseteq \Sigma^*)[B \leq^p_{s\text{-}ld\text{-}T} A \wedge B \not\leq^p_{lnd\text{-}T} A]$.

**Proof** The proof idea is similar to that of Theorem 4.7. We want to construct an oracle $A$ and a test language $B$ such that the membership of each string in $B$ can be tested by asking queries to $A$ in a strong length-decreasing fashion. In particular, the Turing machine implementing the strong query-decreasing reduction from $B$ to $A$ would, on strings of the form $0^{4^k}$, ask $(1/4)4^k$ strong length-decreasing queries to its oracle. The machine then concatenates the answers from the oracle, pads the concatenated string appropriately, and uses the string $q$ thus formed as the next query to its oracle. Our construction ensures that the length of $q$ is less than the lengths of strings queried earlier. Roughly and intuitively speaking, since the bits of $q$ depend on the membership (in $A$) of several strings, a machine cannot compute the value of $q$ without asking to its oracle those queries (of lengths longer than $|q|$) whose membership (in $A$) determine the bits of $q$. This allows us to diagonalize against machines implementing query-nondecreasing reductions to $A$.

For each $D$, we define $L_D$ as follows:

$$L_D = \{0^{4^k} \mid k \geq 1 \wedge 0^{(1/4)4^k} \chi_D(0^{4^k-1}) \chi_D(0^{4^k-2}) \ldots \chi_D(0^{(3/4)4^k}) \in D\}.$$

We will construct sets $A$ and $L_A$ (which is our $B$) such that $L_A \leq^p_{s\text{-}ld\text{-}T} A$, yet $L_A \not\leq^p_{lnd\text{-}T} A$.

Note that, for each $D \subseteq \Sigma^*$, $L_D \leq^p_{s\text{-}ld\text{-}T} D$ via a Turing machine $M$ that rejects any input that is not in $\{0^{4^k} \mid k \geq 1\}$ and that, for any input in $\{0^{4^k} \mid k \geq 1\}$, asks its oracle the following queries in order: $0^{4^k-1}, 0^{4^k-2}, \ldots, 0^{(3/4)4^k}$. Let the answers to these queries be $b_1, b_2, \ldots, b_{(1/4)4^k}$, respectively. $M$ computes the string $q = 0^{(1/4)4^k} b_1 b_2 \ldots b_{(1/4)4^k}$, asks query $q$ to its oracle, and accepts exactly if the oracle says "yes." Since $|q| = (1/4)4^k + (1/4)4^k = (1/2)4^k < (3/4)4^k$, it follows that $M$ indeed has the strong query-decreasing property with respect to each oracle $D$.

We will now describe our construction of $A$. We will construct $A$ in stages. In Stage $i \geq 1$, we will diagonalize against the machine $M_i$. (Recall from Section 2 that for each $X \subseteq \Sigma^*$ and each $x \in \Sigma^*$, $M_i^X(x)$ runs in time $|x|^i + i$.) That is, in Stage $i$ we will show that $M_i$ does not implement $L_A \leq^p_{lnd\text{-}T} A$.

Let $A_{i-1}$ denote the set of strings added to $A$ before Stage $i$. Let $A_0 = \emptyset$.

**Stage $i$, $i \geq 1$:** Let $n = n_i$ be the smallest integer such that, for some $k \geq 1$, $n = 4^k$, $n^i + i < 2^{n/4}$, and no string of length $(1/2)4^k$ or longer has been made relevant (see footnote 6) in any earlier stage.

We will diagonalize against $M_i$. Let $M = M_i$. Let $t$ be the smallest step at which $M^{A_{i-1}}(0^{4^k})$ asks a query of length strictly greater than $(1/2)4^k$. Let $\gamma$ be this string. By definition, $|\gamma| > (1/2)4^k$. (If $M^{A_{i-1}}(0^{4^k})$ never asks a query of length strictly greater than $(1/2)4^k$, let $t = n^i + i$ and let $\gamma$ be undefined.) Let $\alpha$ be the lexicographically smallest string of length $(1/4)4^k$ such that $M^{A_{i-1}}(0^{4^k})$ does not ask the query $0^{(1/4)4^k}\alpha$ in the first $t$ steps. Since $t \leq n^i + i < 2^{(1/4)4^k}$, $\alpha$ is well defined. We will fix the membership (in $A$) of the following strings in accordance with $\alpha$: $0^{4^k-1}$, $0^{4^k-2}$,

..., $0^{(3/4)4^k}$. In particular, for each $1 \le i \le (1/4)4^k$, if the $i$th bit of $\alpha$ is 0, we fix the string $0^{4^k-i}$ to be out of $A$ and if the $i$th bit of $\alpha$ is 1, we fix the string $0^{4^k-i}$ to be in $A$. Formally, let $A' = A_{i-1} \cup \{0^{4^k-i} \mid 1 \le i \le (1/4)4^k \text{ and the } i\text{th bit of } \alpha \text{ is } 1\}$. If $M^{A_{k-1}}(0^{4^k})$ never asks a query a string of length strictly greater than $(1/2)4^k$ (that is, if $\gamma$ is undefined), $M^{A'}(0^{4^k})$ will ask exactly the same set of queries as $M^{A_{i-1}}(0^{4^k})$. Let $A_i = A' \cup \{0^{(1/4)4^k}\alpha \mid M^{A'}(0^{4^k}) \text{ rejects}\}$. $M^{A_i}(0^{4^k})$ accepts if and only if $0^{4^k} \notin L_{A_i}$. We will preserve this behavior. Go to Stage $i+1$.

Otherwise (that is, if $M^{A_{k-1}}(0^{4^k})$ asks a query of length strictly greater than $(1/2)4^k$), $\gamma$ is defined. For each query $s$ that is asked in the first $t-1$ steps of $M^{A_{i-1}}(0^{4^k})$, the membership of $s$ in $A'$ is identical to the membership of $s$ in $A_{i-1}$. Thus, the first $t-1$ steps of $M^{A'}(0^{4^k})$ will be identical to the first $t-1$ steps of $M^{A_{i-1}}(0^{4^k})$. Thus, the query asked by $M^{A'}(0^{4^k})$ at step $t$ will be $\gamma$ and the queries asked in the first $t$ steps of $M^{A'}(0^{4^k})$ will not include $0^{(1/4)4^k}\alpha$. If $M^{A'}(0^{4^k})$ asks the query $0^{(1/4)4^k}\alpha$ at step $t+1$ or later, we know that $M$ does not have the query-nondecreasing property with respect to $A'$ on input $0^{4^k}$ because it asks a query (namely, $0^{(1/4)4^k}\alpha$) of length $(1/2)4^k$ after it asks a query (namely, $\gamma$) of length greater than $(1/2)4^k$. Let $A_i = A'$. $M$ does not implement $L_{A_i} \le^p_{lnd\text{-}T} A_i$ (because $M$ does not have the query-decreasing property with respect to $A'$). We will maintain this behavior throughout the construction. Go to Stage $i+1$.

The remaining case is: $M^{A'}(0^{4^k})$ does not ask the query $0^{(1/4)4^k}\alpha$. Let $A_i = A' \cup \{0^{(1/4)4^k}\alpha \mid M^{A'}(0^{4^k}) \text{ rejects}\}$. By the definition of $L_{A_i}$, $0^{4^k} \in L_{A_i}$ if and only if $M^{A'}(0^{4^k})$ rejects. We will maintain this behavior throughout the construction. Thus, $M$ does not implement $L_{A_i} \le^p_{lnd\text{-}T} A_i$. Go to Stage $i+1$.

**End of Stage $i$.**

Let $A = \bigcup_{i \ge 1} A_i$. It is clear from the construction at Stage $i$ that $M_i$ does not implement $L_A \le^p_{lnd\text{-}T} A$. ❑

As an immediate corollary, we have the following corollary.

**Corollary 4.10**  1. $(\exists A, B \subseteq \Sigma^*)[B \le^p_{s\text{-}ld\text{-}T} A \wedge B \not\le^p_{li\text{-}T} A]$.

2. $(\exists A, B \subseteq \Sigma^*)[B \le^p_{ld\text{-}T} A \wedge B \not\le^p_{li\text{-}T} A]$.

3. $(\exists A, B \subseteq \Sigma^*)[B \le^p_{ld\text{-}T} A \wedge B \not\le^p_{lnd\text{-}T} A]$.

Theorem 4.9 shows that query-nondecreasing Turing reductions are not stronger than strong query-decreasing Turing reductions. Thus, it is natural to ask whether the not-stronger-than relationship holds between query-nonincreasing Turing reductions and strong query-increasing Turing reductions. In Theorem 4.11 we show that this is indeed the case. In fact, the proof of Theorem 4.11 is very similar to that of Theorem 4.9—the difference, roughly, is that in the proof of Theorem 4.9 our construction ensures that the only way a polynomial-time Turing machine can correctly accept/reject the diagonalizing string is by asking queries to the oracle in a length-decreasing fashion, while in the proof of Theorem 4.11 our construction ensures that the only way a Turing machine can correctly accept/reject the diagonalizing string is by asking queries to the oracle in a length-increasing fashion. For the sake of completeness, we give the full proof of Theorem 4.11, but we encourage the reader to work out the details of the proof on his or her own.

**Theorem 4.11** $(\exists A, B \subseteq \Sigma^*)[B \le^p_{s\text{-}li\text{-}T} A \wedge B \not\le^p_{lni\text{-}T} A]$.

**Proof** The proof idea is similar to that of Theorem 4.9. We want to construct an oracle $A$ and a test language $B$ such that the membership of each string in $B$ can be tested by asking queries to $A$ in a strong length-increasing fashion. In particular, the Turing machine implementing the strong query-increasing reduction from $B$ to $A$ would, on strings of the form $0^{4^k}$, ask $(1/4)4^k$ strong length-increasing queries to its oracle. The machine would then concatenate the answers from the oracle, pad the concatenated string appropriately, and use the string $q$ thus formed as the next query to its oracle. The padding ensures that the length of $q$ is greater than the lengths of strings queried earlier. Roughly speaking, since the bits of $q$ depend the membership (in $A$) of several strings, a machine cannot compute the value of $q$ without asking to $A$ those queries (shorter than $|q|$) whose membership (in $A$) determine the bits of $q$. This allows us to diagonalize against machines implementing query-nonincreasing reductions to $A$.

For each $D$, we define $L_D$ as follows:

$$L_D = \{0^{4^k} \mid k \geq 1 \wedge 0^{(5/4)4^k} \chi_D(0^{4^k+1}) \chi_D(0^{4^k+2}) \ldots \chi_D(0^{(5/4)4^k}) \in D\}.$$

We will construct sets $A$ and $L_A$ (which is our $B$) such that $L_A \leq^p_{s\text{-}li\text{-}T} A$, yet $L_A \not\leq^p_{lni\text{-}T} A$.

Note that, for each $D \subseteq \Sigma^*$, $L_D \leq^p_{s\text{-}li\text{-}T} D$ via a Turing machine $M$ that rejects any input that is not in $\{0^{4^k} \mid k \geq 1\}$ and that, for any input in $\{0^{4^k} \mid k \geq 1\}$, asks to its oracle the following queries in order: $0^{4^k+1}$, $0^{4^k+2}$, ..., $0^{(5/4)4^k}$. Let the answers to these queries be $b_1$, $b_2$, ..., $b_{(1/4)4^k}$. $M$ computes the string $q = 0^{(5/4)4^k} b_1 b_2 \ldots b_{(1/4)4^k}$, asks $q$ to its oracle, and accepts exactly if the oracle says "yes." Since $|q| = (5/4)4^k + (1/4)4^k = (3/2)4^k > (5/4)4^k$, it follows that $M$ indeed has the strong query-increasing property with respect to each oracle $D$.

We will now describe our construction of $A$. We will construct $A$ in stages. In Stage $i \geq 1$, we will diagonalize against the machine $M_i$. (Recall from Section 2 that for each $X \subseteq \Sigma^*$ and each $x \in \Sigma^*$, $M_i^X(x)$ runs in time $|x|^i + i$.) That is, in Stage $i$ we will show that $M_i$ does not implement $L_A \leq^p_{lni\text{-}T} A$.

Let $A_{i-1}$ denote the set of strings added to $A$ before Stage $i$. Let $A_0 = \emptyset$.

**Stage $i$, $i \geq 1$**: Let $n = n_i$ be the smallest integer such that, for some $k \geq 1$, $n = 4^k$, $n^i + i < 2^{(1/4)n}$, and no string of length $4^k$ or longer has been made relevant (see footnote 6) in any earlier stage.

We will diagonalize against $M_i$. Let $M = M_i$. Let $t$ be the smallest step at which $M^{A_{i-1}}(0^{4^k})$ asks some query of length strictly less than $(3/2)4^k$. Let $\gamma$ be this string. By definition, $|\gamma| < (3/2)4^k$. (If $M^{A_{i-1}}(0^{4^k})$ never asks a query of length strictly less than $(3/2)4^k$, let $t = n^i + i$ and let $\gamma$ be undefined.) Let $\alpha$ be the lexicographically smallest string of length $(1/4)4^k$ such that $M^{A_{i-1}}(0^{4^k})$ does not ask the query $0^{(5/4)4^k} \alpha$ in the first $t$ steps. Since $t \leq n^i + i < 2^{(1/4)4^k}$, $\alpha$ is well defined. We will fix the membership (in $A$) of the following strings in accordance with $\alpha$: $0^{4^k+1}$, $0^{4^k+2}$, ..., $0^{(5/4)4^k}$. In particular, for each $1 \leq i \leq (1/4)4^k$, if the $i$th bit of $\alpha$ is 0, we put the string $0^{4^k+i}$ out of $A$ and if the $i$th bit of $\alpha$ is 1, we put the string $0^{4^k+i}$ in $A$. Formally, let $A' = A_{i-1} \cup \{0^{4^k+i} \mid 1 \leq i \leq (1/4)4^k$ and the $i$th bit of $\alpha$ is $1\}$. If $M^{A_{i-1}}(0^{4^k})$ never asks a query of length strictly less than $(3/2)4^k$ (that is, if $\gamma$ is undefined), $M^{A'}(0^{4^k})$ will ask exactly the same set of queries as $M^{A_{i-1}}(0^{4^k})$. Let $A_i = A' \cup \{0^{(5/4)4^k} \alpha \mid M^{A'}(0^{4^k})$ rejects$\}$. $M^{A_i}(0^{4^k})$ will accept if and only if $0^{4^k} \notin L_{A_i}$. We will preserve this behavior throughout the construction. Go to Stage $i+1$.

Otherwise (that is, if $M^{A_{i-1}}(0^{4^k})$ asks some query of length strictly less than $(3/2)4^k$), $\gamma$ is defined. The membership (in $A'$) of queries asked in the first $t-1$ steps of $M^{A_{i-1}}(0^{4^k})$ is identical to their membership in $A_{i-1}$. Thus, the first $t-1$ steps of $M^{A'}(0^{4^k})$ will be identical to the first $t-1$ steps of $M^{A_{i-1}}(0^{4^k})$. Thus, the query asked at step $t$ will be $\gamma$ and the queries asked in the first $t$ steps of $M^{A'}(0^{4^k})$ do not include $0^{(5/4)4^k} \alpha$. If $M^{A'}(0^{4^k})$ asks the query $0^{(5/4)4^k} \alpha$ at step $t+1$ or later, we know that $M$ does not have the query-nonincreasing property with respect to $A'$ on input $0^{4^k}$ because it asks a query (namely, $0^{(5/4)4^k} \alpha$) of length $(3/2)4^k$ after it asks a query (namely, $\gamma$) of length less than $(3/2)4^k$. Let $A_i = A'$. $M$ does not implement $L_{A_i} \leq^p_{lni\text{-}T} A_i$. We will maintain this behavior throughout the construction. Go to Stage $i+1$.

The remaining case is: $M^{A'}(0^{4^k})$ does not ask the query $0^{(5/4)4^k}\alpha$. Let $A_i = A' \cup \{0^{(5/4)4^k}\alpha \mid M^{A'}(0^{4^k}) \text{ rejects}\}$. By the definition of $L_{A_i}$, $0^{4^k} \in L_{A_i}$ if and only if $M^{A'}(0^{4^k})$ rejects. We will maintain this behavior throughout the construction. Thus, $M$ does not implement $L_{A_i} \leq^p_{lni\text{-}T} A_i$. Go to Stage $i+1$.
**End of Stage $i$**.

Let $A = \bigcup_{i \geq 1} A_i$. It is clear from the construction at Stage $i$ that $M_i$ does not implement $L_A \leq^p_{lni\text{-}T} A$. ❑

As an immediate corollary, we have the following corollary.

**Corollary 4.12**  1. $(\exists A, B \subseteq \Sigma^*)[B \leq^p_{s\text{-}li\text{-}T} A \wedge B \not\leq^p_{ld\text{-}T} A]$.

2. $(\exists A, B \subseteq \Sigma^*)[B \leq^p_{li\text{-}T} A \wedge B \not\leq^p_{ld\text{-}T} A]$.

3. $(\exists A, B \subseteq \Sigma^*)[B \leq^p_{li\text{-}T} A \wedge B \not\leq^p_{lni\text{-}T} A]$.

What is the relationship between strong query-nonincreasing and query-decreasing reductions? First note that both these reductions are stronger than strong query-decreasing reductions. However, these two reductions seem incomparable: On the one hand strong query-nonincreasing reductions seem more restrictive than query-decreasing reductions because the length of each query in a strong query-nonincreasing reduction must be at least the length of the input. On the other hand, query-decreasing reductions seem more restrictive than strong query-nonincreasing reductions because each query in a query-decreasing reduction must be *strictly* shorter than the previous query. In Theorem 4.13, we show that strong query-nonincreasing Turing reductions and query-decreasing Turing reductions are incomparable. Similarly, we show in Theorem 4.14 that strong query-nondecreasing Turing reductions and query-increasing Turing reductions are incomparable.

**Theorem 4.13** $(\exists A, B, C \subseteq \Sigma^*)[B \leq^p_{ld\text{-}T} A \wedge B \not\leq^p_{s\text{-}lni\text{-}T} A \wedge C \leq^p_{s\text{-}lni\text{-}T} A \wedge C \not\leq^p_{ld\text{-}T} A]$.

**Proof** The proof is based on the construction of an oracle $A$ and two test languages $B = L^1_A$ and $C = L^2_A$ such that (i) $L^1_A \leq^p_{ld\text{-}T} A \wedge L^1_A \not\leq^p_{s\text{-}lni\text{-}T} A$ and (ii) $L^2_A \leq^p_{s\text{-}lni\text{-}T} A \wedge L^2_A \not\leq^p_{ld\text{-}T} A$. To achieve this, we interweave two diagonalizations—the first diagonalization ensures that (i) holds and the second diagonalization ensures that (ii) holds.

The first diagonalization is based on the observation that, on an input of length $n$, the first query in a query-decreasing reduction may legally be of length $n+1$ but in a strong query-nonincreasing reduction the first query cannot be of length $n+1$. For each $D \subseteq \Sigma^*$, define $L^1_D$ as:

$$L^1_D = \{0^{4i} \mid i \geq 1 \wedge 0^{4i+1} \in D\}.$$

Clearly, for each $D \subseteq \Sigma^*$, $L^1_D \leq^p_{ld\text{-}T} D$. (In fact, for each $D \subseteq \Sigma^*$, $L^1_D \leq^p_{\hat{m}} D$, where $\leq^p_{\hat{m}}$ is the variant of $\leq^p_m$ that allows the reduction to when it wishes to directly output "accept" or "reject" [Amb87]. In contrast, $L^1_D \leq^p_m D$ does not hold for the case $D = \Sigma^*$.) Our construction of $A$ will ensure that $L^1_A \not\leq^p_{s\text{-}lni\text{-}T} A$.

The second diagonalization is based on the observation that, on an input of length $n$, asking multiple queries of length $n-1$ is legal in a strong query-nonincreasing reduction but is illegal in a query-decreasing reduction. For each $D \subseteq \Sigma^*$, define $L^2_D$ as:

$$L^2_D = \{0^{4i+3} \mid i \geq 1 \wedge ((0^{4i+2} \in D) \oplus (1^{4i+2} \in D))\}.$$

Clearly, for each $D \subseteq \Sigma^*$, $L^2_D \leq^p_{s\text{-}lni\text{-}T} D$. (In fact, for each $D \subseteq \Sigma^*$, $L^2_D \leq^p_{2\text{-}tt} D$.) Our construction of $A$ will ensure that $L^2_A \not\leq^p_{ld\text{-}T} A$.

We will carry out our interwoven diagonalization in stages. In odd numbered stages, we will carry out the first diagonalization, and in even numbered stages we will carry out the second diagonalization. That is, in odd numbered stages, we will ensure that $L_A^1 \not\leq_{s\text{-}lni\text{-}T}^p A$, while in even numbered stages we will ensure that $L_A^2 \not\leq_{ld\text{-}T}^p A$.

For each $i \geq 1$, let $A_{i-1}$ denote the set of strings in $A$ before the $i$th stage. Let $A_0 = \emptyset$.

**Stage $i = 2j - 1$, $j \in \mathbb{N}^+$.** We will ensure that $M_i$ does not implement $L_A^1 \leq_{s\text{-}lni\text{-}T}^p A$. Let $n = n_i$ be the smallest integer such that no string of length $n$ or longer has been made relevant (see footnote 6) in any previous stage and, for some $k \geq 1$, $n = 4k$. If $M^{A_{i-1}}(0^{4k})$ queries $0^{4k+1}$, let $A_i = A_{i-1}$. $M$ does not have the strong length-nonincreasing property with respect to $A_i$ on input $0^{4k}$. We will maintain this behavior throughout the construction. Go to Stage $i + 1$.

The remaining case is: $M^{A_{i-1}}(0^{4k})$ does not query $0^{4k+1}$. Let $A_i = A_{i-1} \cup \{0^{4k+1} \mid M^{A_{i-1}}(0^{4k})$ rejects$\}$. Note that $M^{A_i}(0^{4k})$ accepts if and only if $0^{4k} \notin L_{A_i}^1$. We will maintain this behavior throughout the construction. Go to Stage $i + 1$.

**End of Stage $i$.**

**Stage $i = 2j$, $j \in \mathbb{N}^+$.** We will ensure that $M_i$ does not implement $L_A^2 \leq_{ld\text{-}T}^p A$. Let $n = n_i$ be the smallest integer such that no string of length $n$ or longer has been made relevant (see footnote 6) in any previous stage and, for some $k \geq 1$, $n = 4k + 3$. If $M^{A_{i-1}}(0^{4k+3})$ queries both $0^{4k+2}$ and $1^{4k+2}$, let $A_i = A_{i-1}$. $M$ does not have the length-decreasing property with respect to $A_i$ on input $0^{4k+3}$. We will maintain this behavior throughout the construction. Go to Stage $i + 1$.

The remaining case is: $M^{A_{i-1}}(0^{4k+3})$ queries at most one string in $\{0^{4k+2}, 1^{4k+2}\}$. If $M^{A_{i-1}}(0^{4k+3})$ accepts, let $A_i = A_{i-1}$. If $M^{A_{i-1}}(0^{4k+3})$ rejects, let $A_i = A_{i-1} \cup \{\alpha\}$, where $\alpha$ is the lexicographically smallest string in $\{0^{4k+2}, 1^{4k+2}\}$ that is not queried by $M^{A_{i-1}}(0^{4k+3})$. $0^{4k+3} \in L_{A_i}$, yet $M^{A_i}(0^{4k+3})$ rejects. We will maintain this behavior throughout the construction. Go to Stage $i + 1$.

**End of Stage $i$.**

Let $A = \bigcup_{i \geq 1} A_i$. It is clear from the construction that $L_A^1 \leq_{ld\text{-}T}^p A$ yet $L_A^1 \not\leq_{s\text{-}lni\text{-}T}^p A$, and that $L_A^2 \leq_{s\text{-}lni\text{-}T}^p A$ yet $L_A^2 \not\leq_{ld\text{-}T}^p A$. ❑

Theorem 4.13 shows that query-decreasing and and strong query-nonincreasing reductions are incomparable. We next show that query-increasing and strong query-nondecreasing reductions are also incomparable.

**Theorem 4.14** $(\exists A, B, C \subseteq \Sigma^*)[B \leq_{li\text{-}T}^p A \wedge B \not\leq_{s\text{-}lnd\text{-}T}^p A \wedge C \leq_{s\text{-}lnd\text{-}T}^p A \wedge C \not\leq_{li\text{-}T}^p A]$

**Proof** The proof is based on the construction of an oracle $A$ and two test languages $B = L_A^1$ and $C = L_A^2$ that simultaneously ensure that (i) $L_A^1 \leq_{li\text{-}T}^p A$ yet $L_A^1 \not\leq_{s\text{-}lnd\text{-}T}^p A$, and (ii) $L_A^2 \leq_{s\text{-}lnd\text{-}T}^p A$ yet $L_A^2 \not\leq_{li\text{-}T}^p A$. To achieve this, we interweave two diagonalizations—the first diagonalization ensures that (i) holds and the second diagonalization ensures that (ii) holds.

The first diagonalization is based on the observation that, on input of length $n$, the first query in a query-increasing reduction may legally be of length $n - 1$ but in a strong query-nondecreasing reduction the first query cannot be of length $n - 1$. For each $D \subseteq \Sigma^*$, define $L_D^1$ as:

$$L_D^1 = \{0^{4i+3} \mid i \geq 1 \wedge 0^{4i+2} \in D\}.$$

Clearly, for each $D \subseteq \Sigma^*$, $L_D^1 \leq_{li\text{-}T}^p D$. (In fact, for each $D \subseteq \Sigma^*$, $L_D^1 \leq_{\hat{m}}^p D$.) Our construction of $A$ will ensure that $L_A^1 \not\leq_{s\text{-}lnd\text{-}T}^p A$.

The second diagonalization is based on the observation that, on an input of length $n$, asking multiple queries of length $n + 1$ is legal in a strong query-nondecreasing reduction but is illegal in a query-increasing reduction. For each $D \subseteq \Sigma^*$, define $L_D^2$ as:

$$L_D^2 = \{0^{4i} \mid i \geq 1 \wedge ((0^{4i+1} \in D) \oplus (1^{4i+1} \in D))\}.$$

Clearly, for each $D \subseteq \Sigma^*$, $L_D^2 \leq_{s\text{-}lnd\text{-}T}^p D$. (In fact, for each $D \subseteq \Sigma^*$, $L_D^2 \leq_{2\text{-}tt}^p D$.) Our construction of $A$ will ensure that $L_A^2 \not\leq_{li\text{-}T}^p A$.

We will carry out our interwoven diagonalization in stages. In odd numbered stages, we will carry out the first diagonalization, and in even numbered stages we will carry out the second diagonalization. That is, in odd numbered stages, we will ensure that $L_A^1 \not\leq_{s\text{-}lnd\text{-}T}^p A$, while in even numbered stages we will ensure that $L_A^2 \not\leq_{li\text{-}T}^p A$.

For each $i \geq 1$, let $A_{i-1}$ denote the set of strings in $A$ before the $i$th stage. Let $A_0 = \emptyset$.

**Stage $i = 2j - 1$, $j \in \mathbb{N}^+$.** We will ensure that $M_i$ does not implement $L_A^1 \leq_{s\text{-}lnd\text{-}T}^p A$. Let $n = n_i$ be the smallest integer such that no string of length $n$ or longer has been made relevant (see footnote 6) in any previous stage and such that, for some $k \geq 1$, $n = 4k+3$. If $M^{A_{i-1}}(0^{4k+3})$ queries $0^{4k+2}$, let $A_i = A_{i-1}$. $M$ does not have the strong length-nondecreasing property with respect to $A_i$ on input $0^{4k+3}$. We will maintain this behavior throughout the construction. Go to Stage $i+1$.

The remaining case is: $M^{A_{i-1}}(0^{4k+3})$ does not query $0^{4k+2}$. Let $A_i = A_{i-1} \cup \{0^{4k+2} \mid M^{A_{i-1}}(0^{4k+3}) \text{ rejects}\}$. Note that $M^{A_i}(0^{4k+3})$ accepts if and only if $0^{4k+3} \notin L_{A_i}^1$. We will maintain this behavior throughout the construction. Go to Stage $i+1$.

**End of Stage $i$.**

**Stage $i = 2j$, $j \in \mathbb{N}^+$.** We will ensure that $M_i$ does not implement $L_A^2 \leq_{li\text{-}T}^p A$. Let $n = n_i$ be the smallest integer such that no string of length $n$ or longer has been made relevant (see footnote 6) in any previous stage and such that, for some $k \geq 1$, $n = 4k$. If $M^{A_{i-1}}(0^{4k})$ queries both $0^{4k+1}$ and $1^{4k+1}$, let $A_i = A_{i-1}$. $M$ does not have the length-decreasing property with respect to $A_i$ on input $0^{4k}$. We will maintain this behavior throughout the construction. Go to Stage $i+1$.

The remaining case is: $M^{A_{i-1}}(0^{4k})$ queries at most one string in $\{0^{4k+1}, 1^{4k+1}\}$. If $M^{A_{i-1}}(0^{4k})$ accepts, let $A_i = A_{i-1}$. If $M^{A_{i-1}}(0^{4k})$ rejects, let $A_i = A_{i-1} \cup \{\alpha\}$, where $\alpha$ is the lexicographically smallest string in $\{0^{4k+1}, 1^{4k+1}\}$ that is not queried by $M^{A_{i-1}}(0^{4k})$. $0^{4k} \in L_{A_i}^2$, yet $M^{A_i}(0^{4k})$ rejects. We will maintain this behavior throughout the construction. Go to Stage $i+1$.

**End of Stage $i$.**

Let $A = \bigcup_{i \geq 1} A_i$. It is clear from the construction that $L_A^1 \leq_{li\text{-}T}^p A$ yet $L_A^1 \not\leq_{s\text{-}lnd\text{-}T}^p A$, and that $L_A^2 \leq_{s\text{-}lnd\text{-}T}^p A$ yet $L_A^2 \leq_{li\text{-}T}^p A$. ❏

The results of this section indicate that unless query-monotonic reductions are "obviously" related, they are provably incomparable. However, thus far while comparing reductions we have used a rather stringent notion, namely the stronger-than relationship. The reason this notion is stringent is that, for $\leq_\beta^p$ to be stronger than $\leq_\alpha^p$, we require that, for *every* pair of sets $A$ and $B$, if $A \leq_\alpha^p B$ then $A \leq_\beta^p B$. In complexity theory, we are often interested in properties that hold for restricted classes of sets with "nice" properties even though these might not hold for all sets. Thus, it is natural to ask how the query-monotonic reductions compare when we restrict ourselves to natural complexity classes (having nice properties). We show in Theorem 4.15 that for each class $\mathcal{C}$ that is closed downward under many-one reductions, the query-nondecreasing (respectively, query-nonincreasing) Turing reduction closure of $\mathcal{C}$ is identical to the Turing reduction closure of $\mathcal{C}$. Since most natural complexity classes have this property, it follows that when the database (oracle) is restricted to be from a natural complexity class, query nondecreasing and query nonincreasing requirements are toothless. (In Sections 5 and 6, we study query-monotonic reductions over NP, arguably the most important complexity class.)

**Theorem 4.15** *For each class $\mathcal{C}$ of languages such that $\mathcal{C}$ is closed downward under $\leq_m^p$ reductions, $\mathrm{R}_T^p(\mathcal{C}) = \mathrm{R}_{lni\text{-}T}^p(\mathcal{C}) = \mathrm{R}_{lnd\text{-}T}^p(\mathcal{C})$*

**Proof** Clearly, for any class $\mathcal{C}$, $\mathrm{R}_{lnd\text{-}T}^p(\mathcal{C}) \subseteq \mathrm{R}_T^p(\mathcal{C})$ and $\mathrm{R}_{lni\text{-}T}^p(\mathcal{C}) \subseteq \mathrm{R}_T^p(\mathcal{C})$, since any query-nondecreasing or query-nonincreasing Turing reduction is also a Turing reduction. Let $A \subseteq \Sigma^*$ and $B \in \mathcal{C}$ be such that $A \leq_T^p B$ via DPTM $M$. If $B = \Sigma^*$, then $A \in \mathrm{P}$, and thus $A \in \mathrm{R}_{lni\text{-}T}^p(\mathcal{C}) \cap$ $\mathrm{R}^p_{lnd\text{-}T}(\mathcal{C})$. So assume that $B \neq \Sigma^*$. We will show that there exists $Z \subseteq \Sigma^*$ and DPTM $M_1$ such that the following properties hold: (a) $Z \leq^p_m B$, (b) $L(M_1^Z) = A$, and (c) for each $x \in \Sigma^*$, all queries that $M_1^Z(x)$ asks to its oracle are of the same length. Then, it follows that $A \in \mathrm{R}^p_{lnd\text{-}T}(Z) \cap \mathrm{R}^p_{lni\text{-}T}(Z)$, and since $\mathcal{C}$ is closed downward under many-one reductions, $Z \in \mathcal{C}$. Thus, it follows that $A \in \mathrm{R}^p_{lnd\text{-}T}(\mathcal{C}) \cap \mathrm{R}^p_{lni\text{-}T}(\mathcal{C})$.

We now specify $Z$ and $M_1$ and show that the properties (a), (b), and (c) mentioned above hold. Roughly speaking, we encode the membership (in $B$) of all strings of length at most $k$ in the membership (in $Z$) of the first $2^{k+1} - 1$ strings of length $k+1$. We achieve this by including the $j$th string of length $k+1$ in $Z$ if and only if $B$ contains the $j$th string in $(\Sigma^*)^{\leq k}$. Now, a sequence of queries to $B$ such that the length of each query is at most $\ell$ can be transformed to a sequence of queries to $Z$ such that the length of each query is *exactly* $\ell + 1$. Formally, $Z$ is defined as follows: $Z = \{x \in \Sigma^* \mid |x| > 0 \ \wedge \ (\exists r : 1 \leq r \leq 2^{|x|-1})[x \text{ is the lexicographically } r\text{th string at length } |x| \text{ and the lexicographically } r\text{th string (in } \Sigma^*) \text{ is in } B]\}$.

Let $a$ be a string that is in $\overline{B}$. (Note that $a$ is well defined because $B \neq \Sigma^*$.) $Z \leq^p_m B$ via the following FP function $\sigma$, which on any input $x \in \Sigma^*$ does the following:

> Compute the rank $r$ of $x$ among strings of length $|x|$. By definition, $1 \leq r \leq 2^{|x|}$. If $r = 2^{|x|}$, output $a$. Otherwise, output the string (in $\Sigma^*$) with lexicographic rank $r$.

Thus, (a) holds. (We mention in passing that, clearly, $B \leq^p_m Z$.) We will now define $M_1$. Let $q$ be a monotonically increasing polynomial that robustly (i.e., for all oracles) bounds the running time of $M$. By definition, such a polynomial exists. For each $x \in \Sigma^*$, $M_1$ on input $x$ does the following. It simulates $M(x)$, and when $M$ asks a query $z$ to its oracle, $M_1$ asks the query $z'$ to its oracle, where $z'$ is the lexicographically $rank(z)$th string at length $q(|x|) + 1$. (Recall from Section 2 that $rank(z)$ is the lexicographic rank of $z$.) Clearly, since $|z| \leq q(|x|)$, $z'$ is well defined. Also, note that $z \in B$ if and only if $z' \in Z$. Thus, the sequence of answers to queries that $M^B(x)$ asks to its oracle is exactly the same as the sequence of answers to queries that $M_1^Z(x)$ asks to its oracle. Also, by construction, all queries that $M_1^Z$ asks to its oracle are of length $q(|x|) + 1$. So, (b) and (c) hold. ❑

As an immediate corollary to Theorem 4.15, we have that for many natural complexity classes (for example, BPP, $\mathrm{C}_=\mathrm{P}$, $\oplus\mathrm{P}$, PP, and each level of the polynomial hierarchy) query-nondecreasing and query-nonincreasing reductions are no less powerful than Turing reductions.

**Corollary 4.16** *For each class* $\mathcal{C} \in \{\mathrm{NP}, \mathrm{NP}^{\mathrm{NP}}, \ldots, \mathrm{coNP}, \mathrm{coNP}^{\mathrm{NP}}, \ldots, \mathrm{P}, \mathrm{P}^{\mathrm{NP}}, \mathrm{P}^{\mathrm{NP}^{\mathrm{NP}}}, \ldots, \mathrm{BPP}, \mathrm{C}_=\mathrm{P}, \oplus\mathrm{P}, \mathrm{PP}\}$, $\mathrm{R}^p_T(\mathcal{C}) = \mathrm{R}^p_{lni\text{-}T}(\mathcal{C}) = \mathrm{R}^p_{lnd\text{-}T}(\mathcal{C})$.

## 5 Query-Monotonic Reductions to NP Sets

In the previous section, we saw that query-monotonic and Turing reductions are different in general, though for many natural complexity classes the reduction closure of these classes with respect to query-nonincreasing, query-nondecreasing, and Turing reductions are identical. In this section, we study the relationship between query-monotonic and Turing reductions to NP sets. We ask the following question: For each $A \subseteq \Sigma^*$ and $B \in \mathrm{NP}$, does $A \leq^p_T B$ imply $A \leq^p_{li\text{-}T} B$? Since $\leq^p_T$ is stronger than $\leq^p_{li\text{-}T}$, if the answer to the above question is "yes," then for each $A \subseteq \Sigma^*$ and for $B \in \mathrm{NP}$, $A \leq^p_T B$ if and only if $A \leq^p_{li\text{-}T} B$. Similarly, we ask the following question: For each $A \subseteq \Sigma^*$ and each $B \in \mathrm{NP}$, does $A \leq^p_T B$ imply $A \leq^p_{ld\text{-}T} B$? Since $\leq^p_T$ is stronger than $\leq^p_{ld\text{-}T}$, if the answer to the above question is "yes," then for each $A \subseteq \Sigma^*$ and $B \in \mathrm{NP}$, $A \leq^p_T B$ if and only if $A \leq^p_{ld\text{-}T} B$. Clearly, if either question has the answer "no," then $\mathrm{P} \neq \mathrm{NP}$. In fact, both these questions are currently open. Indeed, even seemingly easier issues are still open. In particular, is it true that for all $A \subseteq \Sigma^*$ and $B \in \mathrm{NP}$, $A \leq^p_{2\text{-}tt} B$ implies $A \leq^p_{li\text{-}T} B$? In Theorem 5.1 we show that a "yes" resolution of the open questions mentioned above will require nonrelativizable techniques. In fact, we prove that there is an oracle relative to which truth-table (even 2-fixed-truth-table) and query-increasing (query-decreasing) reductions differ on some set in NP. Finally, note that both $\leq^p_{li\text{-}T}$ and $\leq^p_{ld\text{-}T}$ are trivially stronger than $\leq^p_{1\text{-}tt}$ and both $\leq^p_{lni\text{-}T}$ and $\leq^p_{lnd\text{-}T}$ are trivially stronger than $\leq^p_{tt}$, and that both these facts relativize. Thus, the $\leq^p_{2\text{-}tt}$ in Theorem 5.1 cannot be strengthened to $\leq^p_{1\text{-}tt}$ and the $\leq^p_{li\text{-}T}$ (respectively, $\leq^p_{ld\text{-}T}$) in Theorem 5.1 cannot be strengthened to $\leq^p_{lnd\text{-}T}$ ($\leq^p_{lni\text{-}T}$, respectively).

**Theorem 5.1** $(\exists W \subseteq \Sigma^*)(\exists A \in \mathrm{NP}^W)(\exists B \subseteq \Sigma^*)[B \leq^{p,W}_{2\text{-}tt} A \wedge B \not\leq^{p,W}_{li\text{-}T} A \wedge B \not\leq^{p,W}_{ld\text{-}T} A]$.

**Proof** The basic idea of the proof is to exploit the fact that a Turing machine implementing a query-increasing or a query-decreasing reduction cannot ask two queries of the same length. On the other hand a machine implementing a 2-truth-table reduction does not have this restriction. With this insight, we diagonalize against machines implementing query-increasing or query-decreasing reductions by making the membership of the diagonalizing string dependent on the memberships of two strings of the same length in the oracle.

For each $D$, define $L_D$ as follows:

$$L_D = \{0^i \mid i \geq 1 \wedge ((0^i \in D) \oplus (1^i \in D))\}.$$

Clearly, for each $D$, $L_D \leq^{p,W}_{2\text{-}tt} D$. We will construct sets $W$, $A$, and $B = L_A$, such that $A \in \mathrm{NP}^W$, $L_A \leq^{p,W}_{2\text{-}tt} A$, yet $L_A \not\leq^{p,W}_{li\text{-}T} A$ and $L_A \not\leq^{p,W}_{ld\text{-}T} A$. We define $A$ as follows:

$$A = \{0^n \mid n \geq 1 \wedge (\exists y \in \Sigma^*)[|y| = n \wedge 0^n y \in W]\} \cup \{1^n \mid n \geq 1 \wedge (\exists y \in \Sigma^*)[|y| = n \wedge 1^n y \in W]\}.$$

Clearly, $A \in \mathrm{NP}^W$. Note that $L_A$ is implicitly defined in terms of $W$ since $L_A$ is defined in terms of $A$ and $A$ is defined in terms of $W$. We will construct $W$ such that $L_A \not\leq^{p,W}_{li\text{-}T} A$ and $L_A \not\leq^{p,W}_{ld\text{-}T} A$. Recall from Section 2 that $M_1, M_2, \ldots$ is an enumeration of (deterministic) Turing machines such that $M_k$ robustly (that is, for all oracles) runs in time $n^k + k$.

The construction of $W$ will proceed in stages. In Stage $k \in \mathbb{N}^+$, we will diagonalize against machine $M_k$. That is, we will ensure in Stage $k$ that $M_k$ does not implement $L_A \leq^{p,W}_{li\text{-}T} A$ and $M_k$ does not implement $L_A \leq^{p,W}_{ld\text{-}T} A$.

For each $k$, let $W_k$ be the set of all elements added to $W$ before Stage $k$. For each $k$, let $A_k$ be defined as follows:

$$A_k = \{x \in \Sigma^* \mid (\exists y \in \Sigma^*)[|y| = |x| \wedge xy \in W_k]\}.$$

Let $W_1 = \emptyset$.
**Stage $k \geq 1$.** Let $M = M_i$. Choose $n = n_k$ to be smallest positive integer such that $n$ is greater than the length of each string made relevant (see footnote 6) in any earlier stage and $2^n > n^k + k$.

Either $M^{A_k,W_k}(0^n)$ accepts or it rejects. If $M^{A_k,W_k}(0^n)$ accepts, then let $W_{k+1} = W_k$. $M^{A_{k+1},W_{k+1}}(0^n)$ accepts, yet $0^n \notin L_{A_{k+1}}$. (Note: Here we are using the fact that $L_{A_{k+1}}$ is defined in terms of $W_{k+1}$.) Also, our construction will in fact ensure that both these behaviors are preserved, i.e., our construction will ensure that $M^{A,W}(0^n)$ accepts and $0^n \notin L_A$. (We will explain at the end of this proof why these behaviors are preserved.) Otherwise, i.e., if $M^{A_k,W_k}(0^n)$ rejects, then let $Q$ be the set of queries made by $M$ to its first oracle in $M^{A_k,W_k}(0^n)$. Exactly one of the following cases hold.

1. $\|Q \cap \{0^n, 1^n\}\| = 0$.
2. $\|Q \cap \{0^n, 1^n\}\| = 1$.

3. $\|Q \cap \{0^n, 1^n\}\| = 2$.

**Case 1.** Let $W_{k+1} = W_k \cup \{0^n y\}$, where $y$ is the lexicographically smallest string in $\Sigma^n$ such that $M^{A_k, W_k}(0^n)$ does not ask the query $0^n y$ to either of its oracles. Note that such a $y$ exists since $M^{A_k, W_k}(0^n)$ can ask at most $n^k + k$ queries and, by our choice of $n$, $2^n > n^k + k$. Clearly, $0^n \in A_{k+1}$ and $1^n \notin A_{k+1}$, thus $0^n \in L_{A_{k+1}}$. However, $M^{A_{k+1}, W_{k+1}}(0^n)$ rejects. These behaviors will be preserved. Go to Stage $k+1$.

**Case 2.** Consider the case that $0^n \in Q$. (The case when $1^n \in Q$ is analogous.) Let $W_{k+1} = W_k \cup \{1^n y\}$, where $y$ is the lexicographically smallest string in $\Sigma^n$ such that $M^{A_k, W_k}(0^n)$ does not ask the query $1^n y$ to either of its oracles. Note that such a $y$ exists since $M$ can ask at most $n^k + k$ queries and, by our choice of $n$, $2^n > n^k + k$. Clearly, $0^n \notin A_{k+1}$ and $1^n \in A_{k+1}$, thus $0^n \in L_{A_{k+1}}$. However, $M^{A_{k+1}, W_{k+1}}(0^n)$ rejects. These behaviors will be preserved. Go to Stage $k+1$.

**Case 3.** Let $W_{k+1} = W_k$. By definition, $A_{k+1} = A_k$. Since $M^{A_{k+1}, W_{k+1}}(0^n)$ asks two queries, namely $0^n$ and $1^n$, of the same length, it follows that, relative to $W_{k+1}$, $M$ has neither the query-increasing property nor the query-decreasing property with respect to $A_{k+1}$. These behaviors will be preserved. Go to Stage $k+1$.

**End of Stage $k$.**

Let $W = \bigcup_{k \geq 1} W_k$. (Note that $A$ and $L_A$ are above defined in terms of $W$.) Consider an arbitrary $k \geq 1$. Let $M = M_k$. It is clear from the construction at Stage $k$ that at least one of the following conditions holds.

1. $0^{n_k} \in L_{A_{k+1}} \iff 0^{n_k} \notin L(M^{A_{k+1}, W_{k+1}})$.

2. Relative to $W_{k+1}$, $M$ has neither the query-increasing property nor the query-decreasing property with respect to $A_{k+1}$ on input $0^{n_k}$.

Let $k \in \mathbb{N}^+$. It is clear from the construction at Stage $k$ that all the strings added to $A$ in Stage $k$ are from the set $\{0^{n_k}, 1^{n_k}\}$. Since our choice of $n_\ell$, $\ell > k$, is such that $n_\ell$ is greater than the length of any string queried at Stage $k$, the following hold:

1. $0^{n_k} \in L_{A_{k+1}} \iff 0^{n_k} \in L_A$.

2. $0^{n_k} \in L(M^{A_{k+1}, W_{k+1}}) \iff 0^{n_k} \in L(M^{A, W})$.

Thus, for each $k \geq 1$, at least one of the following conditions holds.

1. $0^{n_k} \in L_A \iff 0^{n_k} \notin L(M^{A, W})$.

2. Relative to $W$, $M$ has neither the query-increasing property nor the query-decreasing property with respect to $A$ on input $0^{n_k}$.

Thus, $L_A \not\leq^{p,W}_{li\text{-}T} A$ and $L_A \not\leq^{p,W}_{ld\text{-}T} A$. ❑

Theorem 5.1 shows that proving that query-increasing (query-decreasing) Turing reductions are stronger than truth-table reductions on sets in NP will require nonrelativizable techniques. Thus, it is interesting to ask what (if any) structural consequences follow from the assumption that query-increasing Turing reductions are stronger than truth-table reductions on NP sets. We consider a stronger hypothesis (which we call Hypothesis S), namely that query-increasing Turing reductions are stronger than Turing reductions. Note that under this assumption, for each $A \subseteq \Sigma^*$ and $B \in \text{NP}$, $A \leq^p_T B$ if and only if $A \leq^p_{li\text{-}T} B$. Theorem 5.2 shows that there is an infinite set $L \subseteq \mathbb{N}$ such that if Turing and query-increasing reductions coincide for each set in NP, then for each set $A$ in $\text{P}^{\text{NP}}$, $A \cap \{x \in \Sigma^* \,|\, |x| \in L\}$, the set of all strings in $A$ whose length is in $L$, 1-truth-table reduces to some set in NP. (It is easy to see from the proof that the result holds not only for the particular $L$ mentioned in the proof, but also for each easily recognizable, wide-spaced subset of $\mathbb{N}$.)

**Hypothesis S:** $(\forall A \subseteq \Sigma^*)(\forall B \in \text{NP})[A \leq_T^p B \iff A \leq_{li\text{-}T}^p B]$.

**Theorem 5.2** *There exists an infinite polynomial-time computable set $L \subseteq \mathbb{N}$ such that if Hypothesis S holds then, for each set $A$ in $\text{P}^{\text{NP}}$, there is a $C \in \text{NP}$ such that $A \cap \{x \in \Sigma^* \,|\, |x| \in L\} \leq_{1\text{-}tt}^p C$.*

**Proof** Let $L = \{2^{2^2}, 2^{2^{2^{2^2}}}, 2^{2^{2^{2^{2^{2^2}}}}}, \ldots\}$.

We first informally describe the main idea of the proof. Let $A \in \text{P}^B$ via DPTM $M$, where $B \in \text{NP}$. Let $p$ be the monotonically increasing polynomial $p$ bounding the running time of $M$ independent of the oracle. We will construct a set $C \in \text{NP}$ such that $A \cap \{x \in \Sigma^* \,|\, |x| \in L\} \leq_T^p C$. This set $C$ will be wide-spaced, and in fact, the spacing in $C$ will depend on the spacing in $L$. If Hypothesis S holds, then $A \cap \{x \,|\, |x| \in L\} \leq_{li\text{-}T}^p C$. That is, there is a polynomial-time Turing machine that on any input asks queries of $C$ in a length-increasing fashion and (correctly) accepts $A \cap \{x \,|\, |x| \in L\}$. Note that since $C$ is widely-spaced, there are exponential gaps between lengths at which $C$ potentially contains strings. Let us call the lengths at which $C$ potentially contains strings "interesting." Any polynomial-time machine that, with $C$ as its oracle, asks at most one query per length, asks at most one query at a length that is both large (with respect to the input string's length) and interesting. The membership (in $C$) of any *other* query string can be easily determined *without* querying $C$ in the following manner. First compute whether the query string $q$ is of an interesting length and, if so, brute-force compute the membership of $q$ in $C$. Since $C$ is in NP, and $q$ (due to the *other* above) is exponentially smaller than the input length, this brute-force computation can be done in time polynomial in the length of the input.

Let $L' = \{p(m) + 1 \,|\, m \in L\}$. $C$ will be nonempty only at lengths in $L'$. Define $C$ as: $C = \{x \in \Sigma^* \,|\, |x| \in L' \wedge (\exists r : 1 \leq r \leq 2^{|x|} - 1)[\text{the lexicographic rank of } x \text{ among strings of length } |x| \text{ is } r$ and the lexicographically $r$th string in $\Sigma^*$ is in $B$]$\}$. Roughly speaking, at length $p(m) + 1$, where $m \in L$, $C$ encodes the membership (in $B$) of all strings of length at most $p(m)$.

**Claim 5.3**  *1. $C \in \text{NP}$.*

*2. $A \cap \{x \in \Sigma^* \,|\, |x| \in L\} \leq_T^p C$.*

**Proof** Part 1 follows from the construction of $C$ and the fact that $B \in \text{NP}$. To see that part 2 holds, consider a Turing machine $M_1$ that on input $x$ rejects if $|x| \notin L$. If $|x| \in L$, $M_1(x)$ simulates $M(x)$ and whenever $M$ asks a query $q$ to its oracle, $M_1$ asks a query $q'$ to its oracle, where $q'$ is a string of length $p(|x|) + 1$ such that the lexicographic rank of $q'$ among all strings of length $p(|x|) + 1$ is exactly the lexicographic rank of $q$ (among strings of $\Sigma^*$). $q'$ is well defined since, by assumption, $|q| \leq p(|x|)$. After asking $q'$ to its oracle and obtaining answer $b$ from its oracle, $M_1$ proceeds with the simulation of $M(x)$ assuming that the answer to query $q$ in the simulation of $M(x)$ is $b$. By the construction of $C$, it follows that for each $x \in \Sigma^*$ such that $|x| \in L$, the sequence of answers that $M_1^C(x)$ obtains from its oracle is identical to the sequence of answers that $M^B(x)$ obtains from its oracle. Thus, for each $x \in \Sigma^*$ such that $|x| \in L$, $M_1^C(x)$ accepts if and only if $M^B(x)$ accepts. Clearly, $L(M_1^C) = \{x \in \Sigma^* \,|\, |x| \in L\} \cap L(M^B) = \{x \in \Sigma^* \,|\, |x| \in L\} \cap A$. ❑

We continue with the proof of Theorem 5.2. Now, if Hypothesis S holds, then by Claim 5.3 part 2 it follows that $A \cap \{x \in \Sigma^* \,|\, |x| \in L\} \leq_{li\text{-}T}^p C$. That is, there is a DPTM $M_0$ such that $L(M_0^C) = A \cap \{x \in \Sigma^* \,|\, |x| \in L\}$ and $M_0$ has the query-increasing property with respect to $C$. Let $q$ be a monotonically increasing polynomial such that $M_0$ runs in time bounded by $q$ regardless of the oracle. We will now construct a Turing machine $M_1$ that accepts (with access to $C$) exactly the same set that $M_0^C$ accepts, but we will ensure that $M_1$ on any input asks at most one query to its oracle. Let $x$ be an arbitrary input to $M_1$. $M_1(x)$ rejects if $|x| \notin L$. Otherwise (that is, if $|x| \in L$), $M$ computes $k$, where $k$ is defined as

$$k = \max\{i \in \mathbb{N} \mid i \in L' \land i \leq q(|x|)\}.$$

Note that $k$ is an upper bound on the length of the largest string in $\{x \in \Sigma^* \mid |x| \in L'\}$ that $M_0$ can query on input $x$. Because $C$ contains only strings whose lengths are in $L'$, any string whose length is not in $L'$ is definitely not in $C$. Note that for each $\ell \in L'$ such that $\ell < k$, it holds that $\ell \leq 1 + p(\log\log\log k) \leq 1 + p(\log\log\log q(|x|))$, which is $\mathcal{O}(\log\log|x|)$. Thus, the membership (in $C$) of any string of length $\ell$ such that $\ell \in L'$ and $\ell < k$ can be determined in polynomial time by brute-forcing the NP algorithm for $C$. $M_1(x)$ simulates $M_0(x)$ and when $M_0(x)$ asks a query $q$ of its oracle, $M_1$ determines whether $|q| \in L'$. If not, then $M_1$ assumes that the answer to the query is "no." If $|q| \in L'$, $M_1$ asks the query $q$ to its oracle only if $|q| = k$. Otherwise (that is, if $|q| \in L'$ and $|q| < k$), $M_1$ brute-force computes the membership of $q$ in $C$. Since $M_0$ has the query-increasing property with respect to $C$, it follows that at most one query $M_0^C(x)$ asks is of length $k$, and thus $M_1^C(x)$ asks at most one query to its oracle. Also, as argued earlier, the brute-force computation of all other strings can be done in polynomial time. Thus, $M_1$ is a DPTM and, for each $x \in \Sigma^*$, $M_1^C(x)$ accepts if and only if $M_0^C(x)$ accepts. So, $A \cap \{x \in \Sigma^* \mid |x| \in L\} \leq_{1\text{-}tt}^p C$ via Turing machine $M_1$. ❑

For each set $X$ and any class $\mathcal{C}$, we say that $X$ is $\mathcal{C}$-*immune* if $X$ is an infinite set and no infinite subset of $X$ is in $\mathcal{C}$ [SB84]. For classes $\mathcal{D}$ and $\mathcal{C}$, we say that $\mathcal{D}$ is $\mathcal{C}$-immune if each infinite set in $\mathcal{D}$ is $\mathcal{C}$-immune. (The literature on immunity is large. As a few examples from the past decade, we mention [Rot99,HH03,GOP+05]—see also the references therein.)

One might be fooled into thinking that Theorem 5.2 is a (non)immunity result. However, it is not since, in some cases, $A \cap \{x \mid |x| \in L\}$ will be empty, for example, if $A = \{x \mid |x| \notin L\}$. Nonetheless, if Hypothesis S holds, then for each set $A \in \mathrm{P}^{\mathrm{NP}}$ such that $A \cap \{x \mid |x| \in L\}$ is infinite, $A \cap \{x \mid |x| \in L\}$ is an infinite $\mathrm{R}_{1\text{-}tt}^p(\mathrm{NP})$ subset of $A$, i.e., $\mathrm{P}^{\mathrm{NP}} \cap \{A \mid A \cap \{x \mid |x| \in L\}$ is infinite$\}$ is not $\mathrm{R}_{1\text{-}tt}^p(\mathrm{NP})$-immune. However, we state as a corollary a weaker but more natural result.

**Definition 5.4** *A set $A$ is* length-supported *if it has strings at all but a finite number of lengths, i.e., $\{n \mid A^{=n} = \emptyset\}$ is finite. Let $\mathrm{LS} = \{A \subseteq \Sigma^* \mid A$ is length-supported$\}$.*

**Corollary 5.5** *If Hypothesis S holds, then $\mathrm{P}^{\mathrm{NP}} \cap \mathrm{LS}$ is not $\mathrm{R}_{1\text{-}tt}^p(\mathrm{NP})$-immune.*

On the other hand, one could if one wanted state a stronger (than Theorem 5.2), yet relatively natural, result. In particular, we say that a set of integers $B \subseteq \mathbb{N}$ is *wide-spaced* if, for each $a, b \in B$ such that $a < b$, it holds that $a \leq \log\log\log b$. Note that the proof idea behind Theorem 5.2 clearly yields: If $A \in \mathrm{P}^{\mathrm{NP}}$, then for any polynomial-time computable, wide-spaced set $B$, there exists a $C \in \mathrm{NP}$ such that $A \cap \{x \in \Sigma^* \mid |x| \in B\} \leq_{1\text{-}tt}^p C$. In particular, if $A \cap \{x \in \Sigma^* \mid |x| \in B\}$ is infinite, then it forms an infinite subset of $A$ belonging to $\mathrm{R}_{1\text{-}tt}^p(\mathrm{NP})$. That is, we have as a corollary to the proof technique of Theorem 5.2 the following result.

**Corollary 5.6** *If Hypothesis S holds, then no $\mathrm{P}^{\mathrm{NP}}$ set that, for some polynomial-time computable, wide-spaced set of lengths, has strings at an infinite number of those lengths is $\mathrm{R}_{1\text{-}tt}^p(\mathrm{NP})$-immune.*

## 6 Tight Padding and NP

In Section 4 we saw that query-monotonic reductions are in general different from Turing reductions, and different query sequence collections (for example, $\rho_{li}$, $\rho_{ld}$, etc.) indeed lead to different notions of query-monotonic reductions. On the flip side, in Theorem 4.15 we saw that query-nondecreasing and query-nonincreasing Turing reductions are no more restrictive than Turing reductions for classes of sets that are closed downward under polynomial-time many-one reductions. In this section, we study query-nonincreasing Turing reductions to NP sets. In particular, we ask the following question: For any set $S \in$ NP, what structural property of $S$ is sufficient to ensure that each language $A$ that Turing reduces to $S$ in fact query-increasing (query-decreasing) Turing reduces to $S$? We show in Theorem 6.1 that when the set being reduced to is tight paddable, query-increasing Turing reductions, query-decreasing Turing reductions, strong query-increasing Turing reductions, and Turing reductions are equivalent. Recall from Definition 2.8 that a tight paddable set is a set that has an easily computable function that preserves membership in the set and has strict lower and upper bounds on its output length vis-a-vis the length of the input.

**Theorem 6.1** *Let $S$ be a tight paddable set. Then, for each $A \subseteq \Sigma^*$, the following are equivalent.*

1. $A \leq^p_T S$.
2. $A \leq^p_{li\text{-}T} S$.
3. $A \leq^p_{ld\text{-}T} S$.
4. $A \leq^p_{s\text{-}li\text{-}T} S$.

**Proof** Note that (2) $\implies$ (1), (3) $\implies$ (1), and (4) $\implies$ (2) follow from definitions. We will first prove that (1) $\implies$ (4). The proof relies on the assumption that $S$ is tight paddable. By Definition 2.8, there exist a polynomial-time computable function $\sigma$ and a $k > 0$ such that, for each $x$, (i) $|x| < |\sigma(x)| \leq |x| + k$ and (ii) $x \in S$ if and only if $\sigma(x) \in S$. Let $A$ be an arbitrary set such that $A \leq^p_T S$. Let $M$ be a DPTM, from our enumeration of deterministic, polynomial-time Turing machines (see Section 2) that implements $A \leq^p_T S$. Let $p$ be the monotonically increasing polynomial associated with $M$ in Section 2 that bounds the running time of $M$ regardless of its oracle. By Section 2 (since when $i \in \mathbb{N}$ and $n \in \mathbb{N}$, $n^i + i > n$), for each $n \in \mathbb{N}$, $p(n) > n$. Consider the Turing machine $T$ that, on input $x$, simulates $M(x)$, keeping track of the length, $\ell$, of the query that $T$ most recently asked to its oracle. $T$ initializes $\ell$ to $p(|x|)$. Note that each query that $M^S(x)$ asks will be of length at most $p(|x|)$. $T(x)$ simulates $M(x)$, except when $M(x)$ asks a query $q$ of its oracle, $T$ computes a query $q'$ as follows. Compute $q_0 = q$, $q_1 = \sigma(q)$, $q_2 = \sigma(\sigma(q))$, …, $q_m = \sigma^m(q)$, where $m$ is such that $|q_{m-1}| \leq \ell$ and $|q_m| > \ell$. ($m$ is well defined, in part due to the fact that $|q_0| \leq \ell$.) Note that since $|q_0| > |q_1| > \ldots > |q_m|$, it follows that $m \leq \ell + 1$. Also, since $|q_{m-1}| \leq \ell$, $|q_m| \leq \ell + k$. Let $q'$, the query that $T$ asks to its oracle, be $q_m$. Update the value of $\ell$ to $|q'|$. Note that the new value of $\ell$ is at most $k$ greater than its old value. $T$ now asks the query $q'$ to its oracle. Note that $q' \in S$ if and only if $q \in S$. Thus, the answer $a$ that $T^S(x)$ obtains on query $q'$ is exactly the same as the one that $M^S(x)$ obtains on query $q$. $T$ then proceeds with the simulation of $M(x)$ assuming that the answer $M(x)$ gets to its query $q$ is $a$.

Let $q'_1, q'_2, \ldots, q'_r$ be the queries that $T(x)$ asks of its oracle. From the description above, it is clear that $|x| < p(|x|) < |q'_1| < |q'_2| < \ldots < |q'_r|$. Clearly, $r \leq p(|x|)$ since $M(x)$ runs for at most $p(|x|)$ steps. Furthermore, $|q'_1| \leq p(|x|) + k$, $|q'_2| \leq p(|x|) + 2k$, …, $|q'_m| \leq p(|x|) + rk \leq p(|x|) + p(|x|)k$. Thus, $T^S$ runs in polynomial time.

We can prove (1) $\implies$ (3) using a simulation similar to the one described above except that in this case we will need to start with padded queries that are sufficiently long and, for each subsequent query, use the tight paddability of $S$ to compute (and ask) equivalent queries that are shorter than the previous ones. ❏

Theorem 6.1 shows that for each tight paddable set, the notions of Turing reductions, query-increasing Turing reductions, query-decreasing Turing reductions, and strong query-increasing

Turing reductions are equivalent. It is interesting to note that, for the same sets, it is not clear whether strong query-decreasing Turing reductions are equivalent to Turing reductions (and, to the other three query-monotonic reductions proved, in Theorem 6.1, to be equivalent on tight paddable sets). Intuitively, the reason is that any Turing machine implementing a strong query-decreasing Turing reduction can ask at most $|x|$ queries on input $|x|$ (one query of each of the following lengths: $0, 1, 2, \ldots, |x| - 1$), but this is not necessarily true for a machine implementing a Turing reduction.

In light of Theorem 6.1, it is interesting to ask whether any interesting sets are tight paddable. We show in Theorem 6.3 that many natural NP-complete sets (such as SAT) are tight Z-paddable (and thus, tight paddable). Thus, Theorem 6.1 together with Theorem 6.3 implies that, for any natural NP-complete set $\mathcal{S}$ listed in Theorem 6.3, a polynomial-time computation that uses $\mathcal{S}$ as a database can, without losing any computational power, even access $\mathcal{S}$ in a query-increasing (query-decreasing) fashion.

In Theorem 6.3, we show that a variety of NP-complete problems are tight Z-paddable (and thus, tight paddable). In the proof of Theorem 6.3, we prove that two example NP-complete problems (namely, 3-SAT and CLIQUE) are tight Z-paddable, and leave the rest as an exercise for the reader. Since tight Z-paddability of a set can depend on how the set is encoded, we first specify how we will encode boolean formulas and graphs. The encoding of boolean formulas will be needed in the specification of instances of 3-SAT and the encoding of graphs will be needed in the specification of of instances of CLIQUE.

We first describe how we encode each 3CNF boolean formula. We will define a function $h_{bool}$ such that, for each 3CNF boolean formula $\phi$, $h_{bool}(\phi)$ is a string representing $\phi$. Let $\phi$ be a 3CNF boolean formula over $n$ variables, $x_1, x_2 \ldots, x_n$. Then, $\phi = \phi_1 \wedge \phi_2 \wedge \ldots \wedge \phi_m$, where, for each $i$ such that $1 \leq i \leq m$, $\phi_i = (y_{i,1} \vee y_{i,2} \vee y_{i,3})$, where each literal $y_{i,j} \in \{x_1, x_2, \ldots, x_n\} \cup \{\overline{x_1}, \overline{x_2}, \ldots, \overline{x_n}\}$. Each (uncomplemented) literal $x_i$ is represented as the string $1s_i$ while each complemented literal $\overline{x_i}$ is represented as the string $0s_i$, where $s_i$ is the $i$th string in lexicographic order. $h_{bool}(\phi)$, the representation of $\phi$ as a string, is defined as follows: $h_{bool}(\phi) = \langle w_{1,1}, w_{1,2}, w_{1,3}, w_{2,1}, w_{2,2}, w_{2,3}, \ldots, w_{m,1}, w_{m,2}, w_{m,3} \rangle$, where $w_{i,j}$ is the string representing the literal $y_{i,j}$. It follows easily—from the representation of literals and the fact that the multiarity pairing function (see Section 2) is such that $|\langle x_1, x_2, \ldots, x_k \rangle|$ is $2(k + |x_1| + |x_2| + \ldots + |x_k|)$—that $|h_{bool}(\phi)|$ is $2 \sum_{1 \leq i \leq n} N_i(1 + |s_i|)$, where $n$ is the number of variables in $\phi$, and $N_i$ is the number of occurrences of $x_i$ (as either $x_i$ or $\overline{x_i}$) in $\phi$.

We now describe how to encode an undirected graph. We will define a function $h_{graph}$ such that, for each undirected graph $G$, $h_{graph}(G)$ is a string representing $G$. Let $G = (V, E)$ be an undirected graph over $n$ vertices $V = \{v_1, v_2, \ldots, v_n\}$. $E = \{e_1, e_2, \ldots, e_m\}$, where each $e_j$ is a pair of vertices from $V$. Let $e_j = (v_k, v_\ell)$. Let $w_j = \langle s_k, s_\ell \rangle$, and $h_{graph}(G) = \langle s_n, w_1, w_2, \ldots, w_m \rangle$, where, for each $j \in \mathbb{N}$, $s_j$ is the $j$th string in lexicographic order. Note that the "$s_n$" in $h_{graph}(G)$ denotes the number of nodes in $G$. It is important to encode this information in addition to the edge-connectivity information because the number of nodes cannot be inferred from the edge-connectivity information. For example, given that the set of edges in a graph is $\{(1, 2)\}$, we do not know whether it is a two node graph or, for some $k \in \mathbb{N}^+$, a $k + 2$ node graph with $k$ isolated vertices. It follows easily—from the representation of edges and the bounds on the output length of the multiarity pairing function— that $|h_{graph}(G)|$ is $2(1 + |s_n| + \sum_{1 \leq i \leq n} N_i(1 + |s_i|))$, where $n$ is the number of vertices in $G$ and $N_i$ is the degree of the $i$th vertex.

**Definition 6.2** *(See [GJ79].)*

1. 3-SAT= $\{x \in \Sigma^* \mid$ there is a satisfiable 3CNF boolean formula $\phi$ such that $x = h_{bool}(\phi)\}$.

2. CLIQUE = $\{\langle x, m \rangle \mid x \in \Sigma^*, m \in \mathbb{N}$, and there exists an undirected graph $G$ such that $h_{graph}(G) = x$ and there is a clique of size at least $m$ in $G\}$.

In Theorem 6.3 we claim the tight Z-paddability of several NP-complete problems. We sketch the proof for tight Z-paddability of two such problems, and leave the rest as easy exercises for the reader.

**Theorem 6.3** *The following problems (see [GJ79] for definitions of these problems) are tight Z-paddable (and thus, tight paddable):*

1. 3-SAT.
2. VERTEX COVER.
3. CLIQUE.
4. 3-COLORABILITY.
5. MONOCHROMATIC TRIANGLE.
6. BIPARTITE SUBGRAPH.
7. MULTIPROCESSING SCHEDULING.
8. CYCLIC ORDERING.
9. QUADRATIC DIOPHANTINE EQUATIONS.
10. MAX 2-SAT.
11. DECISION TREE.

**Proof sketch:** The tight Z-padding function $\sigma$ for 3-SAT on input $x$ does the following. If $x$ is not of the correct syntactic form (that is, if there is no boolean formula $\phi$ such that $x = h_{bool}(\phi)$), then clearly $x \notin$ 3-SAT. Let $\sigma(x) = 10x$. Clearly, $|x| < |\sigma(x)| \leq |x| + 2$. From the definition (see Section 2) of $\langle \cdot, \cdot, \ldots, \cdot \rangle$, it follows that $\sigma(x)$ has no preimage in $\langle \cdot, \cdot, \ldots, \cdot \rangle$. Thus, $\sigma(x) \notin$ 3-SAT. On the other hand, if $x$ is of the correct syntactic form (that is, there is a boolean formula $\phi$ such that $x = h_{bool}(\phi)$), then $\sigma(x)$ is defined as follows. Note that in our representation, any nonempty 3CNF boolean formula must have variable $x_1$. Let $\phi' = \phi \wedge (x_1 \vee \overline{x_1} \vee x_1)$. Clearly, $\phi'$ is satisfiable if and only if $\phi$ is satisfiable. Let $\sigma(x) = h_{bool}(\phi')$. By the definition of $h_{bool}$, $|\sigma(x)| = |x| + 6$. Thus, for each $x \in \Sigma^*$, $|x| < |\sigma(x)| \leq |x| + 6$. Also, $\sigma$ is one-to-one, and thus $\sigma$ is a valid tight Z-padding function for 3-SAT.

We now describe a tight Z-padding function $\sigma$ for CLIQUE. On input $x$, $\sigma$ checks whether $x$ is a valid encoding of a CLIQUE instance. If not, $\sigma(x)$ outputs $10x$. From the definition (see Section 2) of $\langle \cdot, \cdot, \ldots, \cdot \rangle$, it follows that $\sigma(x)$ has no preimage in $\langle \cdot, \cdot, \ldots, \cdot \rangle$. Otherwise (that is, if $x$ is a valid encoding of a CLIQUE instance), there is a $G$ and $k \in \mathbb{N}$ such that such that $\langle h_{graph}(G), k \rangle = x$. Let $G$ have $n$ vertices. Let $G'$ be the graph formed by adding $n+1$ isolated vertices (that is, vertices with degree zero) to $G$. (Note that $G'$ has $2n+1$ vertices.) Let $\sigma(\langle x, k\rangle) = \langle h_{graph}(G'), k \rangle$. By the definition of $h_{graph}$, $|h_{graph}(G)| < |h_{graph}(G')| \leq |h_{graph}(G)| + 2$. Thus, for each $z \in \Sigma^*$, $|z| < |\sigma(z)| \leq |z| + 2$. Again, note that adding $n$ isolated vertices increases the size of the representation of the graph by only a constant number of bits because we do not add any additional edges. Thus, $\sigma$ is indeed a tight Z-padding function for CLIQUE. ❑

Theorem 6.1 and Theorem 6.3 together imply that the that reduction closures of 3-SAT with respect to $\leq^p_{li\text{-}T}$, $\leq^p_{s\text{-}li\text{-}T}$, and $\leq^p_{ld\text{-}T}$ reductions are all exactly $P^{NP}$. However, we do not know whether the reduction closure of 3-SAT with respect to $\leq^p_{s\text{-}ld\text{-}T}$ reductions is identical to $P^{NP}$. In other words, we do not know whether $\leq^p_{s\text{-}ld\text{-}T}$ reductions are more restrictive than Turing reductions.

In Theorem 6.4, we give evidence that it is somewhat unlikely that the $\leq^p_{s\text{-}ld\text{-}T}$ reduction has exactly the same power as the Turing reduction.

**Theorem 6.4** *If* $R^p_T(3\text{-SAT}) = R^p_{s\text{-}ld\text{-}T}(3\text{-SAT})$, *then* $P^{NP} = P^{NP[n]}$.

**Proof** The result follows from the fact that if $A \leq^p_{s\text{-}ld\text{-}T} B$ via a Turing machine $M$, then, for each $x$, $M^B(x)$ asks at most $|x|$ queries to its oracle. ❑

Note that, on input $x$, the base machine implementing a Turing reduction to SAT can easily (i.e., in polynomial time) brute-force strings that represent boolean formulas with at most $\log n$ variables (or even $k \log n$ variables, for any particular, fixed $k$). Thus, for these "small" strings the machine need not make queries to its oracle. By this observation and the fact that any boolean formula having $m$ variables appearing in it must need $\Omega(m \log m)$ bits, Theorem 6.4 can be slightly improved: For each $k \in \mathbb{N}$, even $P^{NP} = P^{NP[n - k \log n \log \log n]}$ can be claimed as the conclusion.

Theorem 6.3 shows that several natural NP-complete problems are tight Z-paddable. Are *all* NP-complete sets tight Z-paddable? What about other NP sets? We prove that each NP set (except $\emptyset$ and $\Sigma^*$) is polynomial-time many-one equivalent to both a tight Z-paddable set and one which is not tight paddable set (and thus, certainly not tight Z-paddable). Since each set that is polynomial-time many-one equivalent to an NP-complete set is itself NP-complete, we obtain as a corollary that there are NP-complete sets that are not tight paddable (and thus, certainly not tight Z-paddable). Similarly, we obtain as a corollary that there are polynomial-time computable sets that are tight Z-paddable (and thus, certainly tight paddable).

**Theorem 6.5** *Each set, except $\emptyset$ and $\Sigma^*$, is polynomial-time many-one equivalent to a set that is not tight paddable.*

**Proof** Let $A$ be an arbitrary set other than $\emptyset$ and $\Sigma^*$. If $A$ is a finite set, let $B$ be the set of all strings whose lengths are perfect squares. That is, let $B = \{x \mid (\exists n \in \mathbb{N})[|x| = n^2]\}$. Otherwise (that is, if $A$ is an infinite set), we will encode (in $B$) at length $n^2$ membership of all strings (in $A$) of length $n$. Formally, $B = \{x \in \Sigma^* \mid (\exists n, r \in \mathbb{N})[n^2 = |x|$, the lexicographic rank of $x$ among all strings of length $n^2$ is $r$, $r \leq 2^n$, and the lexicographically $r$th string of length $n$ is in $A]\}$.

Clearly, $A \leq^p_m B$, and $B \leq^p_m A$ in both cases (i.e., when $A$ is finite and when $A$ is infinite). Now, assume for the purpose of deriving a contradiction that $B$ is tight paddable. Under this assumption there exists a function $\sigma : \Sigma^* \to \Sigma^*$ and $k \in \mathbb{N}$ such that for each $x \in \Sigma^*$, $|x| < |\sigma(x)| < |x| + k$, and $x \in B$ if and only if $\sigma(x) \in B$. If $A$ is an finite set, let $x = 0^{n^2}$, where $n$ is the smallest integer in $\mathbb{N}$ such that $(n+1)^2 > n^2 + k$. Otherwise (that is, if $A$ is an infinite set), let $x$ be a string such that $x \in B$ and $(\lfloor \sqrt{|x|} \rfloor + 1)^2 > |x| + k$. Note that in both cases (that is, when $A$ is finite and when $A$ is infinite), $x$ is well defined. Also, in both cases $x \in B$. Now, $|x| < |\sigma(x)| < |x| + k < (\lfloor \sqrt{|x|} \rfloor + 1)^2$. Since $B$ is only nonempty at lengths that are perfect squares, and the smallest perfect square that is greater than $|x|$ is $(\lfloor \sqrt{|x|} \rfloor + 1)^2$, it follows that $|\sigma(x)|$ is not a perfect square, and thus $\sigma(x)$ is not in $B$, which is a contradiction. ❑

As an immediate corollary, we have the following result.

**Corollary 6.6**
1. *Each set in NP, except $\emptyset$ and $\Sigma^*$, is many-one equivalent to a set that is not tight paddable (and thus, not tight Z-paddable).*

2. *There exists an NP-complete set that is not tight paddable (and thus, not tight Z-paddable).*

3. *There exists a polynomial-time computable set that is not tight paddable (and thus, not tight Z-paddable).*

Turning to the positive side, we have the following result.

**Theorem 6.7** *Every set is polynomial-time many-one equivalent to a tight Z-paddable set.*

**Proof** $\emptyset$ and $\Sigma^*$ are each tight Z-paddable via the function $\sigma(x) = 0x$. Let $A$ be a set other than $\emptyset$ and $\Sigma^*$. We will now define $B$, a tight Z-paddable set such that $A$ is polynomial-time many-one equivalent to $B$. Informally speaking, $B$ at length $n+1$ encodes, in the first $2^{n+1} - 1$ strings, membership (in $A$) of all strings of length at most $n$. Formally, $B = \{x \in \Sigma^* \mid |x| > 0 \land (\exists r \in \mathbb{N})[1 \leq r \leq 2^{|x|} - 1,\ x$ is the $r$th string in lexicographic order among all strings of length $|x|$, and the string with lexicographic rank $r$ (in $\Sigma^*$) is in $A]\}$. Note that $1^n$ is the last string in lexicographic order among all strings of length $n$. By the definition of $B$ above, for each $n$, $1^n \notin B$.

$A \leq_m^p B$ via a function that, roughly speaking, maps each string $x$ to the string $y$ at length $|x|+1$ such that the rank of $y$ among strings of length $|x|+1$ is $rank(x)$. $B \leq_m^p A$ via a function $\sigma'$ that on any input $x \notin \{1^n \mid n \geq 1\}$, outputs the string $y$ such that $rank(y)$ is equal to the lexicographic rank of $x$ among all strings of length $|x|$. For each $n$, $\sigma'(1^n)$ outputs a fixed string that is not in $A$. (Such a string exists because $A \neq \Sigma^*$.) Thus, $A$ and $B$ are polynomial-time many-one equivalent.

Now, consider the function $\sigma : \Sigma^* \to \Sigma^*$ defined in the following manner. For each $n \in \mathbb{N}$, $\sigma(1^n) = 1^{n+1}$. For each string $x$ such that $x \notin \{1^n \mid n \geq 1\}$, let $r$ be the lexicographic rank of $x$ among all strings of length $|x|$. Note that since $x \neq 1^{|x|}$, it follows that $1 \leq r \leq 2^{|x|} - 1$. $\sigma(x)$ outputs $y$, where $y$ is the string of length $n+1$ that has rank $r$ among all strings of length $n+1$. Clearly, $\sigma$ is one-to-one. Note that $x \in B$ if and only if the lexicographically $r$th string (in $\Sigma^*$) is in $A$. Also, $y \in B$ if and only if the lexicographically $r$th string is in $A$. Thus $x \in B$ if and only if $y \in B$. Since $|\sigma(x)| = |x| + 1$, for each $x$, it follows that $\sigma$ is a tight Z-padding function for $B$. ❑

As an immediate corollary, we obtain the following result.

**Corollary 6.8**  1. *Every* NP *set is polynomial-time many-one equivalent to a tight Z-paddable set.*

2. *Every* NP*-complete set is polynomial-time many-one equivalent to a tight Z-paddable set.*

3. *Every polynomial-time computable set is polynomial-time many-one equivalent to a tight Z-paddable set.*

In Theorems 6.1, 6.5, and 6.7, we explored the reducibility properties of tight paddable (and tight Z-paddable) NP sets. It is interesting to ask whether there are some structural properties that are common to all paddable sets in NP. We show that all S-paddable NP sets are *self-witnessing*. Self-witnessing NP (or SelfNP) was studied by Homan and Thakur [HT03b] in their study of one-way permutations and the relationships between an NP set and the set of witnesses (accepting paths of some NPTM) of the strings in the set. SelfNP is the subclass of NP containing all sets $L$ such that, relative to some NPTM, the set of witnesses of strings in $L$ is exactly equal to $L$.

**Definition 6.9** *[HT03b]*

1. *For each* NPTM $N$ *and each* $x \in L(N)$, $y$ *is a witness of* $N$ *on input* $x$ *if* $y$ *is an accepting path in* $N(x)$, *that is,* $y$ *is a sequence of nondeterministic guesses that results in* $N(x)$ *accepting.*

2. *For each* NPTM $N$, $wit_N$ *is a function that maps each* $x \in L$ *to the set of witnesses of* $N$ *on input* $x$.

3. SelfNP $= \{L \mid (\exists$ NPTM N$)[L(N) = L \land \bigcup_{x \in L} wit_N(x) = L]\}$.

Homan and Thakur prove that even though SelfNP and NP are unlikely to be the same, $\text{R}_m^p(\text{SelfNP}) = \text{NP}$. We show in Theorem 6.12 that any S-paddable NP set is self-witnessing (that is, in SelfNP). The proof of this result will use a result due to Homan and Thakur, which we state as Theorem 6.11. Before stating Theorem 6.12, we state some definitions that will be required in the statement of Theorem 6.11 as well as the proof of Theorem 6.12.

**Definition 6.10** *[HT03b]*

1. *An* NPTM $M$ *is* self-contained witnessing *if* $\bigcup_{x \in L(M)} wit_M(x) \subseteq L(M)$.

2. *An* NPTM $M$ *is* honest *if there exists a polynomial $p$ such that for all $x \in L(M)$, there exists a $w \in wit_M(x)$ such that $p(|w|) \geq |x|$.*

Homan and Thakur show that if an NP set is accepted by a self-contained witnessing NPTM, then it is in SelfNP.

**Theorem 6.11** *[HT03b] For each language $L \in$ NP, if there exists a self-contained witnessing, honest NPTM that accepts $L$, then $L \in$ SelfNP.*

We can now show that each S-paddable set (and thus, each Z-paddable set) in NP is self-witnessing.

**Theorem 6.12** *If $L \in$ NP and $L$ is S-paddable (see Definition 2.6), then $L \in$ SelfNP.*

**Proof** Let $L \in$ NP via NPTM $N$ such that $p$ is a monotonically increasing polynomial bounding the length of paths in $N$. Let $L$ be S-paddable. From Proposition 2.7, we can claim w.l.o.g. that $L$ is S-paddable via $\sigma : \Sigma^* \times \Sigma^* \to \Sigma^*$ such that $\sigma$ is invertible in both arguments and, for each $x, y \in \Sigma^*$, $|\sigma(x, y)| > |x|$. Let $q$ be a monotonically increasing polynomial bounding the length of output of $\sigma$. Thus, for each $x, y$, $|\sigma(x, y)| \leq q(|x| + |y|)$. We will prove that $L$ is accepted by a self-contained witnessing, honest NPTM. From Theorem 6.11, it follows that $L \in$ SelfNP.

We will first describe the proof idea informally and then describe it in detail. We construct from $N$ an honest NPTM $N'$ that accepts the same language (namely, $L$) as $N$, but the witnesses of $N'$ are all members of $L$. Suppose $w$ is a witness of $x$ in $N$. Then, our construction will ensure that the corresponding witness of $x$ in $N'$ is $\sigma(x, w)$. Since $x$ has a witness (namely, $w$) in $N$, $x \in L$. Since $\sigma$ is an S-padding function for $L$, $\sigma$ preserves membership in $L$. Thus, $\sigma(x, w) \in L$. Also, $|\sigma(x, w)| > |x|$ ensures that $N'$ is honest.

Let $N'$ be a nondeterministic Turing machine that on any input $x \in \Sigma^*$ does the following:

> Nondeterministically guess a string $w$ of length at most $q(|x| + p(|x|))$. Compute whether $w$ has a preimage in $\sigma$. That is, (deterministically) compute whether there exist $x'$ and $w'$ such that $\sigma(x', w') = w$. (Note that $\sigma$ is polynomial-time invertible in both arguments, so we can deterministically compute the preimage of $w$.) If $w$ has no preimage in $\sigma$, then reject on the current path. Otherwise, let $x'$ and $w'$ be such that $\sigma(x', w') = w$. If $x' \neq x$, then reject on the current path. If $x' = x$, then accept on the current path if and only if $N(x)$ on the path $w'$ accepts.

Let $x \in L$. Then, by definition, there exists a path $w$ of length at most $p(m)$ such that $N(x)$ accepts. By our construction above, $N'(x)$ accepts on the path $\sigma(x, w)$ and $|\sigma(x, w)| > |x|$. On the other hand, if $x \notin L$, then on each path $w$ in $N(x)$, it rejects. By construction, $N'(x)$ rejects on all paths. Thus, $L(N') = L(N) = L$, and $N'$ (since, for all $x$ and $y$, $|\sigma(x, y)| > |x|$) is an honest NPTM. To see that $N'$ is a self-contained NPTM note that, by construction, if $w$ is a witness in $N(x)$, then $w = \sigma(x, w')$, for some $w'$ and $x \in L$. Thus, from the properties of $\sigma$, it follows that $w \in L$. Thus, $N'$ is a self-contained witnessing, honest NPTM and $L(N') = L$. Thus, by Theorem 6.11, $L \in$ SelfNP. ❑

Note that all natural NP-complete problems are known to be S-paddable [BH77]. Thus, it follows from Theorem 6.12 that essentially all natural NP-complete languages are in fact in SelfNP.

# 7 Conclusion and Open Problems

In this paper, we introduced a notion of query-monotonicity in Turing reductions that forces oracle Turing machines to make a rapid progression through their information source. We defined various query-monotonic reductions and compared the power of these reductions to those of Turing reductions. We proved that even though in general these reductions are different (weaker) from Turing reductions, there are classes for which these notions coincide. In particular, we defined a structural property, namely tight paddability, on sets that is sufficient to ensure that query-increasing and query-decreasing Turing reductions coincide with Turing reductions. It would be interesting to find interesting structural conditions on a set that are necessary to ensure that query-increasing and query-decreasing Turing reductions are equivalent to Turing reductions to that set.

Recall that in the introduction we mentioned that the motivation for our study was formalizing the notion of rapid progress through an information source (database/oracle). We mentioned that our reductions have a "use it when you can" flavor: Once one query at or beyond a length is asked, the rest of the information at that length is forever lost to direct access, on the current input. The rapidness of progress in query-increasing and query-decreasing reductions is enforced by the *strict* length increasing (decreasing) nature of those reductions. Similarly, the rapidness of progress in query-nondecreasing (query-nonincreasing) reductions is enforced by each query being required to be at least as long (at most as long) as the previous query. We leave as an open issue the study of a different notion: requiring each query to be lexicographically greater than (less than) the previous query. This notion allows up to a polynomial number of "hits" of the information at each length (in contrast to the one "hit" of query-increasing/ query-decreasing reductions), yet retains a stronger notion of progress though the information source than do query-nondecreasing/query-nonincreasing reductions.

**Acknowledgment** We thank Chris Homan for a helpful conversation and Mitsunori Ogihara for helpful suggestions, both related to the material of Section 6.